\documentclass[12pt]{article}
\usepackage{amsfonts}
\usepackage{float}
\usepackage{jcappub}
\usepackage{graphicx}
\graphicspath{{figures/}}
 \usepackage{caption}
 \usepackage{subcaption}
 \usepackage{epstopdf}
 \usepackage{hyperref}
 \usepackage{cleveref}
 \usepackage{enumitem}
 \usepackage{pdfpages}
 \usepackage{multicol}
\usepackage{mathrsfs}
\usepackage{amssymb}
\usepackage{textcomp}
\usepackage{amsmath, float}
 \usepackage{times}
 \usepackage{etoolbox}
\usepackage{color}
\usepackage[normalem]{ulem}
\usepackage{enumitem}
\usepackage{mathpazo} 

\usepackage[english]{babel}
\usepackage[nottoc]{tocbibind}

\begin{document}

\title{\boldmath  A Non-Parametric Test of Variability of Type Ia Supernovae Luminosity and CDDR }

\author[a]{Darshan Kumar}
\author[b]{, Akshay Rana}
\author[c]{, Deepak Jain}
\author[a]{, Shobhit Mahajan}
\author[a]{, Amitabha Mukherjee}
\author[d,e,f]{, and R. F. L. Holanda}
\affiliation[a]{\small Department of Physics and Astrophysics, University of Delhi, \\Delhi 110007, India}
\affiliation[b]{\small St. Stephen’s College, University of Delhi, \\Delhi 110007, India}
\affiliation[c]{\small Deen Dayal Upadhyaya College, University of Delhi,\\ Dwarka, New Delhi 110078, India}
\affiliation[d]{\small Departamento de Física, Universidade Federal do Rio Grande do Norte, \\Natal-Rio Grande do Norte 59072-970, Brasil}
\affiliation[e] {\small Brasil Departamento de Física, Universidade Federal de Campina Grande, \\ Campina Grande - PB 58429-900, Brasil} 
\affiliation[f] {\small Departamento de Física, Universidade Federal de Sergipe,\\ Aracaju - SE 49100-000, Brazil}
\emailAdd{dkumar1@physics.du.ac.in}
\emailAdd{akshay@ststephens.edu}
\emailAdd{djain@ddu.du.ac.in} 
\emailAdd{sm@physics.du.ac.in}
\emailAdd{am@physics.du.ac.in}
\emailAdd{holandarfl@fisica.ufrn.br}

\abstract{{The first observational evidence for cosmic acceleration appeared from  Type Ia supernovae (SNe Type Ia) Hubble diagram from two different groups. However, the empirical treatment of SNe Type Ia and their ability to show cosmic acceleration have been the subject of some debate in the literature. In this work we probe the assumption of redshift-independent absolute magnitude $(M_{\mathrm{B}})$ of SNe along with its correlation with spatial curvature ($\Omega_{k0}$) and cosmic distance duality relation (CDDR) parameter  ($\eta(z)$). This work is divided into two parts. Firstly, we check the validity of CDDR which relates the luminosity distance ($d_L$) and angular diameter distance ($d_A$) via redshift. We use the Pantheon SNe Ia dataset combined with the $H(z)$ measurements derived from the cosmic chronometers. Further, four different redshift-dependent parametrizations of the distance duality parameter $(\eta(z))$ are used. The CDDR is fairly consistent for almost every parametrization within a  $2\sigma$ confidence level in both  flat and a non-flat universe. In the  second part, we assume the validity of CDDR and emphasize on the variability of $M_{\mathrm{B}}$ and its correlation with $\Omega_{k0}$. We choose four different redshift-dependent parametrizations of $M_{\mathrm{B}}$. The results indicate no evolution of $M_{\mathrm{B}}$ within $2\sigma$ confidence level. For all parametrizations, the best fit value of $\Omega_{k0}$ indicates a flat universe at $2\sigma$ confidence level. However a mild inclination towards a non flat universe is also observed.} We have also  examined the dependence of the results on the choice of different priors for  $H_0$.
\vspace{5mm}\\
{\textbf{Keywords:} supernova type Ia - standard candles, cosmological parameters from LSS, dark energy theory, cosmology of theories beyond the SM.}
}

\maketitle
\flushbottom

\section{Introduction}

The first observational evidence for the cosmic acceleration appeared from  Type Ia supernovae (SNe Type Ia) observations performed by two different research groups \cite{Riess1998,Perlmutter1999}. Over the decades, the number of SNe Type Ia catalogs have increased significantly. Even now, SNe Type Ia observations provide the most direct evidence for the current cosmic acceleration. In the context of Einstein's General Theory of Relativity (GTR), SNe Type Ia observations support the existence of a mysterious form of energy called dark energy, that is either constant ($\Lambda$CDM) or slowly varying with time and space. (See reviews in Refs. \citep{Peebles2003,Caldwell2009,Weinberg2013}).\\

Type Ia supernovae (SNe Type Ia) are extremely luminous  explosions and these are observationally  identified by the absence of hydrogen and presence of silicon (Si II) spectral lines in their spectra \citep{Hillebrandt2000}.  The use of SNe Type Ia as a reliable cosmological probe relies on two fundamental assumptions: 

a) The first assumption is that  the shape of the light curve of all type Ia supernovae (SNe Type Ia) is similar; hence these can be standardized  \citep{phillips1993,Tripp1998}. In practice,  this task of standardizing light curve of SNe Type Ia is achieved by using several empirically derived light curve fitters like, for instance, MLCS/MLCS2k2 \citep{2007ApJ...659..122J}, SALT \citep{Salt2},  SALT2 \citep{Salt2}, SiFTO \citep{Conley_2008}  etc. As stated, the generalized functional forms of  the light curves obtained in these fitters are purely empirical and based on plausible physical explanations to explain the light curves of SNe Type Ia from the time of explosion to a few weeks after peak brightness\citep{Zheng_2017}.

b) The second basic assumption behind the SNe Type Ia analysis is that the intrinsic luminosity of a supernova is independent of the redshift and the host galaxy environment. In other words, it is {theoretically} assumed that two different SNe Type Ia in two different host galaxies have the same intrinsic luminosity, independent of masses and redshifts of the host galaxies. However, in recent years several dedicated studies have concluded that {the light curve fitting analysis of SNe Type Ia  depends on their host galaxy masses \cite{2018ApJ...859..101S}.}  Further, it is seen that  SNe Type Ia events occurring in  massive early-type, passive galaxies are brighter than  those in late-type star forming galaxies \citep{2013ApJ...770..108C,kim_2019,gallagher2008,2000Hamuy}.  

The two main progenitor models of SNe Type Ia are the ``single degenerate'' and the ``double degenerate'' models. In the single degenerate model,  a white dwarf accreting  from a binary companion star is pushed over the Chandrasekhar mass limit, while in the double degenerate model of SNe Type Ia explosions, an orbiting pair of  binary white dwarfs merge together and their mass eventually exceeds the Chandrasekhar mass limit. There is no established formalism to exactly differentiate between these two channels of SNe Type Ia explosion. Hence the observed supernova population catalog may have SNe Type Ia contributions coming from both these channels which will have a direct impact on the first and second assumptions used in SNe Type Ia analysis \citep{wright2018type}.\\ 

Further, the statistical treatment of SNe Type Ia, their dimming and their ability to prove the cosmic acceleration is still a topic of debate in the literature. This is because the process of obtaining the SNe Type Ia observations is not trivial but  requires some priors for interpretations and corrections  to convert an observed-frame magnitude to a rest-frame magnitude. Examples discussed in the literature  are for instance: possible evolutionary effects in SNe Type Ia events \citep{Drell2000,Combes2004}, local Hubble bubble \citep{Zehavi1998,Conley2007}, modified gravity \citep{Ishak2006,Kunz2007,Bertschinger2008},  unclustetered sources of light attenuation \citep{Aguirre1999,Rowan2002,Goober2002,Goober2018} and the existence of Axion-Like-Particles (ALPs) arising in a wide range of well-motivated high-energy physics scenarios.  All these could also lead to the dimming of SNe Type Ia brightness \citep{Avgoustidis2009,Avgoustidis2010} which would eventually affect the second assumption used in the SNe Type Ia analysis. \\

 Given the above mentioned limitations, Tutusaus et al. (2019) \cite{Tutusaus2019} relaxed the standard assumption that SNe Type Ia intrinsic luminosity is independent of the redshift and examined its  impact  on the cosmic acceleration. The authors reconstructed the expansion rate of the Universe by fitting the SNe Type Ia observations with a cubic spline interpolation. They showed  that a non-accelerated expansion rate of the Universe  is able to fit all the main background cosmological probes. In addition, Tutusaus et al. (2017) \cite{Tutusaus2017}  found that when SNe Type Ia intrinsic luminosity is not assumed to be redshift independent, a non-accelerated low-redshift power law model is able to fit the low-redshift background data as well as the $\Lambda$CDM model. It has also been found that a significant correlation exists  between SNe Type Ia luminosity (after the standardization) and the stellar population age at a 99.5\%  confidence level \citep{Kang2020} indicating that the light-curve fitters used by the SNe Type Ia community are not quite capable of correcting for the population age effect. More recently, Valentino et al. (2020) \citep{Valentino_2020} have also raised the issue whether intrinsic SNe Type Ia luminosities might evolve with redshift. They analyse the impact of the latter on the inferred properties of the dark energy component responsible for cosmic acceleration. However, they find the evidence for cosmic acceleration to be robust to possible systematics.  Along the same lines, Sapone et al. (2020) \citep{sapone2020measurable} also analyse the cosmological implications of an absolute luminosity of SNe Type Ia  which could vary with respect to the redshift. Further, they study the impact of the latter on modified gravity models and non-homogeneous models.\\
 
These statements seem to offer enough plausible reasons to investigate the effects of  variability of absolute luminosity of SNe Type Ia. However, these are not the sole reasons of concern about the variability of $M_B$. The cosmic acceleration rate and the cosmological parameters determined by the SNe Type Ia measurements are highly dependent on the possible dimming effect as well. Vavrycuk et al. (2019) \cite{Vav2019} revived a debate about an origin of Type Ia supernova (SN Ia) dimming and showed that the standard $\Lambda$CDM model and the opaque universe model (caused by light extinction by intergalactic dust) fit the SN Ia measurements at redshifts $z < 1.4$ fairly well. Then, there are still some possible loopholes in the  current SNe Type Ia observations and alternative mechanisms contributing to the acceleration evidence or even mimicking the dark energy behaviour have been proposed.  It is important to point out that a constant value of  the absolute luminosity of SNe Type Ia at the peak of its light curve is not sensitive to Hubble constant ($H_0$). Nevertheless,a variable  absolute magnitude  will be sensitive to $H_0$.\\
 
These recent development motivate us to carry out a  model independent study of finding any correlation between  variable luminosity of SNe Type Ia and observed SNe Type Ia dimming effect due to opacity of the environment.  The most general methodology of testing the cosmic opacity of SNe Type Ia is based on the cosmic distance duality relation (CDDR) which connects the luminosity distance $d_L$ and angular diameter distance (ADD) $d_A$ at the same redshift and is defined as, $d_L(1+z)^{-2}/d_A=\eta(z)=1$. In order to look for the presence of some unknown physics phenomenon beyond the standard model or an inconsistency between cosmological data we check if $\eta(z)\neq1$, that is CDDR is violated. This relation holds for general metric theories of gravity in any background, in which photons travel { along} null geodesics and the number of photons is conserved during cosmic evolution \citep{Ellis2007}. Briefly, the SNe Type Ia observations have been confronted with  several other cosmological probes (strong gravitation lens systems, angular diameter distances, gas mass fractions, baryon acoustic oscillations, cosmic microwave background, radio sources, gravitational waves, $H(z)$ measurements, gamma ray bursts etc.) in order to put limits on the redshift dependence of $\eta$, that is  on $\eta(z)$ \cite{Holanda2010,Lima2011,Li2011,Gonalves2011,Meng2012,Holanda2012,Yang2013,Liang2013,Shafieloo2013,Zhang:2014eux,SantosdaCosta2015,Jhingan2014,Chen2015,Holanda2016,Rana2016,Liao2016,Holanda2016b,Holanda2017,Rana2017,Lin2018,Fu2019,Ruan2018,Holanda2019,Chen2020,Zheng2020,Kumar:2020ole,Hu2018}. All these works conclude that CDDR  is valid within a $2 \sigma$ confidence level. However, it is worth stressing  that current analysis could not distinguish which { functional form of} $\eta(z)$  best describes the data (see details in Ref.\cite{William2020}).\\

Another cosmological parameter which can be  crucial to the variability of absolute luminosity of SNe Type Ia and its dimming effect is the cosmic curvature. Cosmic curvature ($\Omega_{k0}$) is a fundamental geometric quantity of the Universe.  It plays a crucial role in the evolution and dynamics of  the universe. As, $\Omega_{k0}$ is directly related to  cosmological distances, a flat or non-flat space-time would obviously impact the path travelled by the photon and eventually the absolute magnitude of SNe Type Ia.  Hence, in this paper, we also probe the variation of the absolute luminosity of SNe Type Ia and its correlation with the  CDDR parameter ($\eta(z)$) and  cosmic curvature ($\Omega_{k0}$).\\

For clarity of exposition, this paper is divided into two parts. In the first part, we test the validity of CDDR. We propose a new  cosmological non-parametric 
test for CDDR by using SNe Type Ia observations and $H(z)$ measurements from cosmic chronometers. The basic procedure is as follows: we obtain the angular diameter distances at SNe Type Ia redshifts by applying Gaussian Process integration method on cosmic chronometer $H(z)$ data. By using a deformed 
CDDR of the form $d_L=\eta(z) d_A(1+z)^2$ 
and considering a flat universe, we impose limits on the SNe Type Ia absolute magnitude ($M_B$), and $\eta(z)$ functions, namely: $\eta(z)=\eta_0, \eta(z)=\eta_0 + \eta_1z$, $ \eta(z)=\eta_0 + \eta_1z/(1+z)$ and $\eta(z)=\eta_0+\eta_1 \text{ln}(1+z)$. \\

After testing the validity of the CDDR parameter $\eta$ in the first part, in the second part, {we assume the validity of CDDR and} put limits on the  possible evolution of  SNe Type Ia absolute magnitude by assuming $M_B(z)= M_{B0}$, $M_B(z)=M_{B0}+M_{B1}z$, $M_B(z)=M_{B0}+M_{B1}z/(1+z)$ and $M_B(z)=M_{B0}+M_{B1}\text{ln}(1+z)$ parametrizations in flat and non flat cosmologies. 
\\

The structure of the paper is as follows: In Section 2, we discuss  the cosmological probes and the data sets used in the analysis along with their theoretical construction. In Section 3, we outline the methodology used.  In Section 4, the  emphasis is on the results of both parts where in the  first part we test the validity of CDDR and in the  second part we test the variability of absolute luminosity of SNe Type Ia. Finally, in Section 5,  we  discuss the final outcomes of both parts and also explore the impact of different $H_0$ priors on our analysis.

\section{Cosmological Probes and Data sets}

In this section we present the cosmological probes and data sets used in the analysis.

\subsection{Type Ia Supernovae measurement and Pantheon Sample} 

Type Ia supernovae are considered to be standard candles. Observations of these are at  the core of establishing the validity of cosmic acceleration. The standard observable quantity in  SNe Type Ia analysis is the distance modulus which is the difference between the apparent and absolute magnitude of the SNe Type Ia. The observational measurement of this quantity is given by the relation

\begin{equation}\label{equ:cddr2}
\mu_{\mathrm{SN}}=m_{\mathrm{B}}^{\mathrm{obs}}(z)+\alpha \cdot X_{1}-\beta \cdot \mathcal{C}-M_{\mathrm{B}}.
\end{equation}
where, $m_{\mathrm{B}}$ is the rest frame B-band observed peak magnitude, $X_{1}$ and $\mathcal{C}$ are time stretching of the light curve and SNe Type Ia color at maximum brightness respectively and $M_{\mathrm{B}}$ is the absolute B-band magnitude. This relation indicates that the variability of the distance modulus is governed by two additional parameters $X_1$ and $\mathcal{C}$. It must be noted here that we do have two nuisance parameters $\alpha$ and $\beta$ as well. In this paper we use the recent and largest database of SNe Type Ia known as the Pantheon data set. In this data set, these two nuisance parameters have been marginalized and eventually calibrated to be zero. This data set has $1048$ SNe Type Ia measurements in the redshift range $0.01\leq z\leq2.26$ \cite{2018ApJ...859..101S}. Hence, for the Pantheon data set, the observed distance modulus relaxes to the form $\mu_{\mathrm{SN}}=m_{\mathrm{B}}^{\mathrm{obs}}-M_{\mathrm{B}}$. Once we know the distance modulus, we can easily define the luminosity distance and uncertainty in observed luminosity distance as

\begin{equation}\label{equ:cddr3}
d_{\mathrm{L}}^{\mathrm{th}}(z;M_{\mathrm{B}})=10^{\left(m_{\mathrm{B}}^{\mathrm{th}}-M_{\mathrm{B}}-25\right) / 5}(\mathrm{Mpc}), 
\end{equation}

From Eq.(\ref{equ:cddr3}), it can be easily seen  that once we estimate the value of $M_{\mathrm{B}}$, we can find the luminosity distance at a given redshift. However, in this work, our aim is to study the variation of the absolute magnitude ($\mathrm{M}_B$) of SNe Type Ia, 
 so if we can get a model-independent estimate of the luminosity distance from other observational probes then we can constrain the variability of $\mathrm{M}_B$.\\

We use the cosmic distance duality relation (CDDR) to reconstruct the luminosity distance theoretically which is given by
{
\begin{equation}\label{equ:cddr3aa}
d_\mathrm{L}^{\mathrm{th}}(z;\eta,\Omega_{k0})=\eta(z) d_\mathrm{A}(z;\Omega_{k0})(1+z)^2.
\end{equation}

Using Eq. (\ref{equ:cddr3aa}) we can define, from Eq. (\ref{equ:cddr3}), $m_{\mathrm{B}}^{\mathrm{th}}$ as

\begin{equation}\label{equ:cddr3bb}
m_{\mathrm{B}}^{\mathrm{th}}(z;\eta,M_{\mathrm{B}},\Omega_{k0})=5\log\left(\eta(z)d_A(z;\Omega_{k0})(1+z)^2\right)+M_{\mathrm{B}}+25, 
\end{equation}
}

{Here, $\eta$ is the cosmic distance duality parameter which is a measure of  the deviation from CDDR. CDDR holds  for $\eta(z)=1$. }
 Here $d_\mathrm{L}^{\mathrm{th}}$ is the theoretical luminosity distance defined in the terms of angular diameter distance  $d_\mathrm{A}$ and $\eta(z)$. The angular diameter distance is defined as 
 
 \begin{equation}
  d_A(z;H_0,\Omega_{k0}) = \begin{cases}
    \dfrac{d_H}{(1+z)\sqrt{\Omega_{k0}}}\sinh\left[\sqrt{\Omega_{k0}} \dfrac{d_C}{d_H}\right] & \mbox{for $\Omega_{k0}>$0 }.\\
    \dfrac{d_C}{(1+z)} & \mbox{for $\Omega_{k0}=$0}.\\
    \dfrac{d_H}{(1+z)\sqrt{|\Omega_{k0}}|}\sin\left[\sqrt{|\Omega_{k0}|} \dfrac{d_C}{d_H}\right] & \mbox{for $\Omega_{k0}<$0}.\\
  \end{cases} \label{rdef}
\end{equation}


Here $\Omega_{k0}$ is the cosmic curvature, where $\Omega_{k0}$ is greater, equal and less than  zero for open, flat  and closed universe respectively. $d_C$ is the comoving distance and  $d_H= c/H_0$ is known as the Hubble distance where $c$ is the speed of light and $H_0$ is the Hubble constant. \\

\subsection{ Constraining Angular Diameter Distance using $H(z)$ measurements}

 As is evident from Eq. \ref{rdef}, if we can obtain a model-independent estimate of  $ d_C/d_H$ then we can easily estimate angular diameter distance $d_\mathrm{A}(z;\Omega_{k0})$ as a function of $\Omega_{k0}$ and subsequently theoretical luminosity distance  $d_\mathrm{L}^{\mathrm{th}}(z;\eta,\Omega_{k0})$ as a function of $\eta$ and $\Omega_{k0}$. In this section  we will first discuss the data sets and  the methodology to obtain angular diameter distances from them. 

\subsubsection{Hubble data set}

In cosmology, the Hubble parameter $H(z)$ is a crucial measured quantity which describes the dynamical properties of the universe such as the expansion rate and evolution history of the universe. It is also helpful to explore the nature of the dark energy. The most recent data compilation of Hubble parameter measurements \citep{2018MNRAS.476.1036M} has $31$ measurements of  $H(z)$  which are obtained by using the differential ages of passively evolving galaxies. We now outline the steps needed to obtain the angular diameter distance that we need.\\

\begin{enumerate}[label=(\roman*)]
	\item  \textbf{Differential ages of passively evolving galaxies:} The Hubble parameter $H(z)$ can be expressed in the terms of the rate of change of cosmic time with the redshift, given by
	
	\begin{equation}
	H(z)= - \dfrac{1}{(1+z)} \dfrac{\Delta z}{\Delta t}.
	\end{equation} 
	The change of cosmic time with redshift can be estimated from the ageing of the stellar population in the galaxies. However, one needs to be extremely careful in  selecting the galaxies while calculating the $H(z)$ values using this method. In young evolving galaxies, the stars are being  born continuously and the emission spectra will be dominated by the young stellar population. Hence to estimate accurately the differential ageing of the universe, passively evolving red galaxies are used as their light is mostly  dominated by the old stellar population \citep{Jimenez_2002}. To find  $H(z)$ at a given redshift, the ages of the early type passively evolving galaxies with similar metallicity and very small redshift interval is calculated. The redshift difference $\Delta z$ between two galaxies can be measured  by using  spectroscopic observations. For the estimation of $\Delta t$, \citep{Moresco_2012} suggested the use of a direct spectroscopic observable (the 4000 Å break) which is known to be linearly related to the age of the stellar population of a galaxy at fixed metallicity. As the measure of $H(z)$ is estimated purely by using spectroscopic observations, it is independent of the cosmological model and has  proved to be a strong probe to constrain cosmological models and assumptions. This method of calculating $H(z)$ is usually known as the ``Cosmic Chronometers'' and data points are generally referred to as CC $H(z)$. We have $31$ data points estimated using this differential ages of passively evolving galaxies technique \citep{Zhang_2014, PhysRevD.71.123001,Stern_2010,Moresco_2012,10.1093/mnrasl/slv037,Moresco_2016}. 
	
		\begin{figure}[]
		\centering
		\includegraphics[height=5cm,width=7.6cm,scale=4]{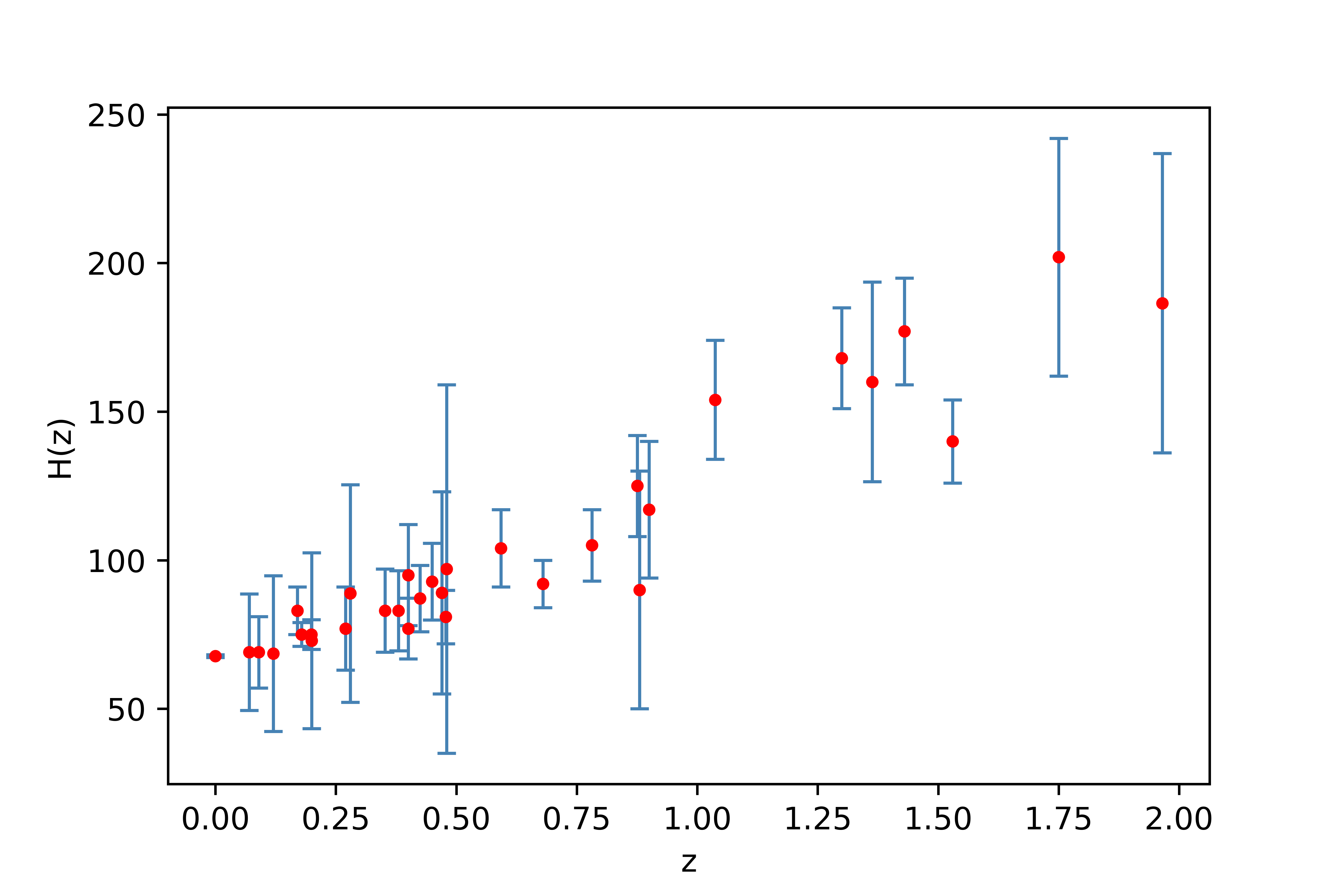}
		\includegraphics[height=5cm,width=7.6cm,scale=4]{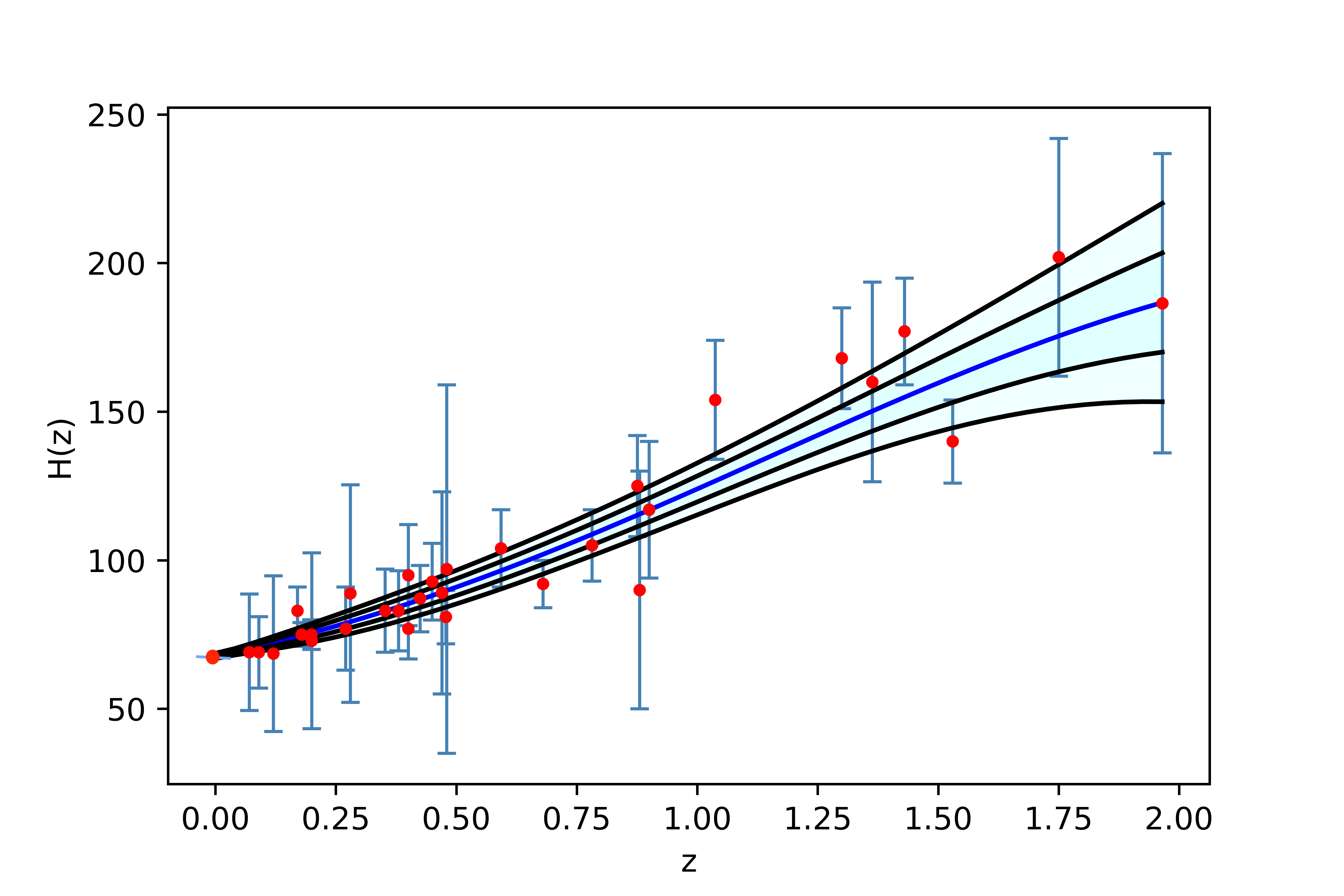}
		\caption{ In the left plot, 31 CC $H(z)$ vs $z$ datapoints are shown.  In the right plot,  $H(z)$ is estimated at all intermediate redshifts in the range $0< z<2$ using a non-parametric smoothening technique, namely Gaussian Process.  In the analysis, the Hubble constant value is taken to be $H_0 = 67.66 \pm 0.42 \mathrm{~km} \mathrm{~s}^{-1} \mathrm{Mpc}^{-1}$ estimated from CMB measurement \cite{Planck2018}.  The impact of other choices of $H_0$ values on analysis has been discussed in Section \ref{discussion}(C).}
		\label{fig:hz31}
	\end{figure}

\end{enumerate}

\subsubsection{Comoving distance using Gaussian Process}

In this analysis, we require a model-independent estimate of the angular diameter distance. For this, we first calculate the comoving distance $d_C$ using the $H(z)$ measurement of cosmic chronometers as

\begin{equation}
    \dfrac{d_C}{d_H} = \int\limits_0^z \dfrac{dz'}{E(z')}.
    \label{dcdh}
\end{equation}

	\begin{figure}[h]
	\centering
	\includegraphics[height=6cm,width=9.5cm,scale=4]{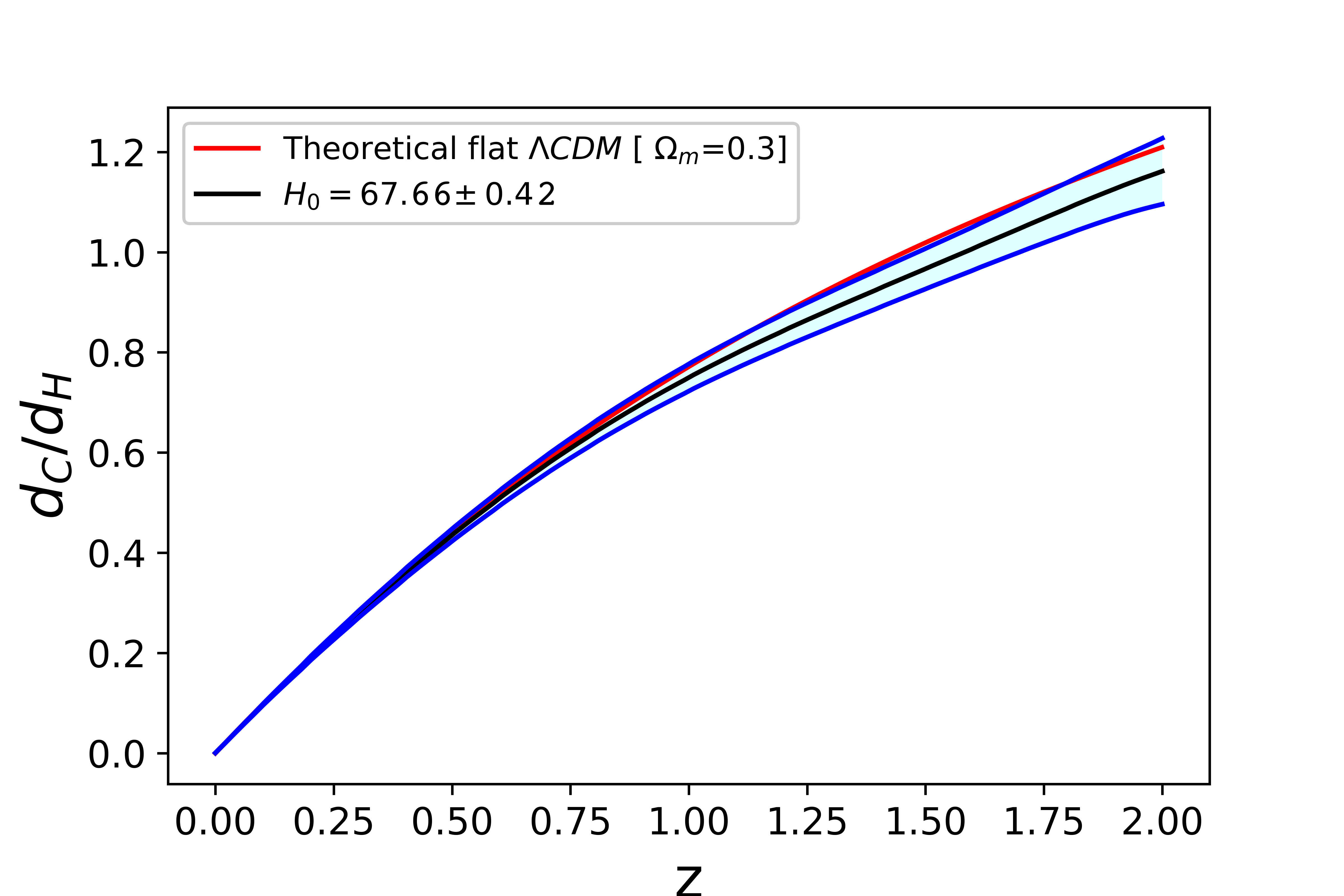}
	\caption{ This plot represents the reconstructed values of $d_C/d_H$ in the redshift in the range $0< z<2$. The solid black line represents the best fit line while blue lines includes the $1 \sigma$ confidence region. The solid red line is the theoretical curve for the flat $\Lambda CDM$ model with $\Omega_m= 0.3$. The value of $H_0$ used in reconstructing this plot is $H_0 = 67.66 \pm 0.42 \mathrm{~km} \mathrm{~s}^{-1} \mathrm{Mpc}^{-1}$.
		\label{fig:hz311}
}
\end{figure}
Here $E(z)= H(z)/H_0$. If we have the functional form of $E(z)$ then we can easily integrate Eq. \ref{dcdh}  by using any numerical integration method. However, in our case we  have only $H(z)$ estimates at certain redshifts. {{In our analysis, to obtain  continuous smooth values of $H(z)$ we use the  Gaussian Process (GP), 
a well known hyper-parametric regression method \cite{2006gpml}. GP has been widely used in the literature to reconstruct the shapes  of  physical functions and is very useful for such functional reconstructions due to its flexibility and simplicity. In this method, the complicated parametric relationship is replaced by parametrizing a probability model over the data. In mathematical terms, it is a distribution over functions, characterized by a mean function and covariance function, given by}}

\begin{equation}
    K\left(z, z^{\prime}\right)=\left\langle(H(z)-\mu(z))\left(H\left(z^{\prime}\right)-\mu\left(z^{\prime}\right)\right)\right\rangle
\end{equation}

{{ where $\mu(z)$ is the prior mean. In order to avoid model dependence appearing through the choice of the  prior mean function, we have chosen it to be zero. This is a common choice of mean function as one can always normalize the data so it has zero mean. We have also checked by taking different values of prior mean, $\mu(z)$ and observed that the result is independent of the choice of the prior mean function.} This method, however, comes with a few inherent underlying assumptions- it is assumed that each observation is an outcome of an independent Gaussian distribution belonging to the same population and the outcomes of observations at any two redshifts are correlated with the strength of correlation depending on their nearness to each other. We reconstructed our data by using the square exponential kernel function, given by}

\begin{equation}
    K(z,z')= \sigma_f^2 \exp{\left(-\dfrac{(|z-z'|)^2}{2\ell^2}\right)}
\end{equation}

{ Here, $\sigma_f$ and $l$ are two hyperparameters which control the amplitude and length-scale of the prior covariance. The value of hyperparameters is calculated by maximizing the corresponding marginal log-likelihood probability function of the distribution. For maximization,  we use  flat priors for $\sigma_f$ and $l$ for all the choices of kernel function. In order to check the sensitivity of our analysis to the choice of the kernel function, we repeated our analysis with the Matérn-(3/2, 5/2, 7/2 \& 9/2) kernel functions. Though the values of hyperparameters vary according to the choice of kernel function, {the reconstructed curves do not show any significant deviation from the curve  obtained from square  exponential curve.} Hence we choose to work with the square exponential kernel function only.}

Once we obtain the reconstructed $H(z)$ at all possible redshifts in the range $0<z<2$ as shown in Figure \ref{fig:hz31}, we divide it by  $H_0 = 67.66 \pm 0.42 \mathrm{~km} \mathrm{~s}^{-1} \mathrm{Mpc}^{-1}$ value to obtain $E(z)$. We use the Simpson $3/8$ method for  numerically integrating  Eq. \ref{dcdh} and obtain continuous values of $d_C/d_H$ at all redshifts in the range $0<z<2$ which is shown in Figure \ref{fig:hz311}. Further, we can use Eq. \ref{rdef} to obtain the angular diameter distance as a function of cosmic curvature $\Omega_{k0}$. The uncertainty in $d_A$ i.e  $\sigma_{d_A}$ is estimated by propagating the error obtained in $d_C/d_H$ using Gaussian Process.


\section{Analysis}
Generally, the Pantheon dataset comes in terms of SNe Type Ia apparent magnitudes and with a full covariance matrix\footnote{\url{http://github.com/dscolnic/Pantheon}}, $C_{\text{sys}}$ correlating the apparent magnitudes at various redshifts. This covariance matrix is a non-diagonal matrix of systematic uncertainties that come from the bias corrections method\cite{2018ApJ...859..101S,2020PhRvD.102b3520K}.

In this analysis, we have to fit simultaneously three parameters i.e. $M_\mathrm{B}$, $\eta$ and $\Omega_{k0}$. These parameters are determined by maximizing the likelihood $\mathcal{L} \sim \exp \left(-\chi^{2} / 2\right)$, where chi-square $(\chi^2)$ is a quantity summed over all the Pantheon SNe Type Ia Sample redshifts, and is defined as

\begin{equation}\label{equ:cddr4aa}
\chi_{\mathrm{Pan}}^{2}=\Delta m^{T} \cdot C^{-1} \cdot \Delta m
\end{equation}
where, $C=D_{\text{stat}}+C_{\text{sys}}$. Here $D_{\text{stat}}$ is the diagonal covariance matrix of the statistical uncertainties and   $\Delta m=m_B^{\mathrm{obs}}(z_i)-m_B^{\mathrm{th}}(z_i;\eta,M_{\mathrm{B}},\Omega_{k0})$ which is given in Eqs.(\ref{equ:cddr2},\ref{equ:cddr3bb}).

In this analysis, we have three unknown parameters, namely, the  absolute magnitude of SNe Type Ia, $M_B$, the cosmic distance duality parameter, $\eta(z)$, and  the cosmic curvature, $\Omega_{k0}$. In order to investigate the variability of $M_B$ we have to analyse its correlation with the remaining two parameters. In order to do so, we divide the work into two parts:

\subsection*{Part I: Test of CDDR}

In this part, we consider the distance duality relation parameter $(\eta)$ to test CDDR and put constraints on $\eta(z)$ simultaneously with the  other two parameters i.e. $M_\mathrm{B}$ and $\Omega_{k0}$. We take into account four parametrizations of $\eta(z)$ 
which are as follows 
\begin{itemize}
    \item P1: $\eta(z) = \eta_0$.
    \item P2: $\eta(z)=\eta_0+\eta_1 z$.
    \item P3: $\eta(z)=\eta_0+\eta_1\dfrac{z}{1+z}$.
    \item P4: $\eta(z)=\eta_0+\eta_1\text{ln}(1+z)$.
\end{itemize}

We included these to study the impact of different characterizations on the analysis. The motivation for this form of  parametrizations comes from the commonly used parametrizations  for the  equation of state parameter, viz. the  Chevallier-Polarski-Linder (CPL) Parametrization, Jassal-Bagla-Padmanabhan (JBP) Parametrization etc \cite{2001IJMPD..10..213C,2003PhRvL..90i1301L,2005MNRAS.356L..11J}.

In the first parametrization, we are choosing $\eta$ to be redshift independent. In second parametrization, it is a simple Taylor series expansion around $z=0$ but is not well behaved at higher z values. The third and fourth parametrizations are taken as these are well behaved even at high redshifts and are somewhat more  slowly varying as compared to the second one.

Further in each parametrization of $\eta(z)$, 
we discuss two cases. In the first case we choose a flat universe by considering $\Omega_{k0}=0$  and then put constraints on $M_\mathrm{B}$ and $\eta$. In the second case, we consider $\Omega_{k0}$ as a free parameter corresponding  to a  non-flat universe.

\subsection*{Part II: Test of variability of SNe Type Ia absolute luminosity }

After testing the validity of CDDR in Part I, we solely focus on the variability of absolute magnitude of SNe Type Ia and use the following parametrizations of $M_B$

\begin{itemize}
    \item M1: $M_B(z) = M_{B0}$.
    \item M2: $M_B(z) = M_{B0}+ M_{B1}  z$.
    \item M3: $M_B(z) = M_{B0}+ M_{B1}\dfrac{z}{1+z}$.
    \item M4: $M_B(z) = M_{B0}+ M_{B1}\text{ln}(1+z)$.
\end{itemize}

In each parametrization of $M_B(z)$, we discuss two cases. In the first case we choose a flat universe and then put constraints on $M_\mathrm{B}$ and in the second case,  we consider a non-flat universe.

We use  $\textbf{{\texttt{emcee}}}$, a Python based package, to perform the Markov Chain Monte Carlo (MCMC) analysis \cite{2013PASP..125..306F}. We find the best fit of all parameters and once the MCMC method is performed, the confidence level with their $1\sigma$, $2\sigma$  and $3\sigma$ uncertainties are computed with the Python package $\textbf{{\texttt{corner}}}$\cite{2016JOSS....1...24F}.

\vspace{3mm}

{In this work, we assume a broad flat prior for all parameters (Table \ref{tabj:1}) . We use 100 walkers which take 10000 steps for  exploring the parameter space with MCMC chains. The first $10\%$ of the steps are discarded as burn-in period and the posterior distributions are analysed based on the remaining samples. To ensure that the chains are converging, an auto-correlation study is performed. For this we compute the integrated auto-correlation time $\tau_f$ using the \textbf{\textit{autocorr.integrated\_time}} function of the \textbf{\textit{emcee}} package. For more details, please see ref.\cite{2013PASP..125..306F} .

}
\begin{table}[h]
	\centering
	\renewcommand{\arraystretch}{2}
	\begin{tabular}[b]{| c | c|}\hline
		Parameter & Prior Range\\ \hline \hline
		$M_\mathrm{B}$ & U[-21,-17]\\ \hline
		$\eta_0$ & U[-3,3] \\ \hline
		$\eta_1$ & U[-2,2] \\ \hline
		$\Omega_{k0}$ & U[-2.5,2.5]\\ \hline
	\end{tabular}
	\caption{ The prior range of $M_\mathrm{B},~\eta_0$, $\eta_1$ and $\Omega_{k0}$.}
	\label{tabj:1}
\end{table}


\section{Results}
In this paper our aim is to probe the variation of the absolute luminosity $(M_B)$ of SNe Type Ia.
This paper is divided into two parts. In the first part, we propose a new (cosmological) model-independent test for  CDDR by using SNe Type Ia observations and $H(z)$ measurements from cosmic chronometers.  In the second part we assume  that  CDDR is valid and check the dependency of absolute magnitude on redshift.

\subsection{Test of CDDR}
In this part, we consider the distance duality parameter $(\eta)$ and take into account four parametrizations of $\eta(z)$. In each parametrization, we discuss two cases- flat and non-flat universe. 

{
\subsubsection*{P1. $\eta(z)=\eta_0$.}

Taking $\eta(z)$ to be a constant, the best fit values of $M_\mathrm{B}$, $\eta_0$ and $\Omega_{k0}$ parameters are given in Table \ref{tb:table4}.

\begin{table}[h]
	\centering
	\renewcommand{\arraystretch}{2}
	\begin{tabular}[b]{| c | c |c|}\hline
		Parameter & Flat Universe & Non-Flat Universe \\ \hline \hline
		$M_\mathrm{B}$ & $-19.300^{ +0.812}_{ -0.874}$ & $-19.653^{ +0.587}_{ -0.431}$ \\ \hline
		$\eta_0$ & $0.960_{-0.476}^{+0.300}$ & $1.124^{ +0.250}_{ -0.265}$ \\ \hline
		$\Omega_{k0}$ & --- & $0.076^{ +0.110}_{ -0.106}$\\ \hline
	\end{tabular}
	\caption{ The best fit values of $M_\mathrm{B},~\eta_0$ and $\Omega_{k0}$ with $1\sigma$ confidence level for P1 parametrization.}
	\label{tb:table4}
\end{table}

{ The best fit value of $\eta(z)$ shown in Table \ref{tb:table4} for both flat as well as  non-flat universe, indicates that the cosmic distance duality relation $(\eta(z)=1)$ holds at $1\sigma$ confidence level. In both flat and non-flat case, the value of $M_\mathrm{B}$ remains same within $1\sigma$ confidence level which reflects that $M_\mathrm{B}$ doesn't have any strong dependence on the curvature parameter. The 1D and 2D posterior distributions of $M_\mathrm{B}, \eta_0$ and $\Omega_{k0}$ with $1\sigma$, $2\sigma$ and $3\sigma$ confidence levels for flat and non-flat universe are shown in Fig.\ref{fig:Part_I_P1}. The contour plots shown in Fig.\ref{fig:Part_I_P1} indicate a negative correlation between $M_B$ and $\eta_0$. Further, this parametrization supports a flat universe within $1\sigma$ confidence level. \\}


\begin{figure}
\centering
\begin{subfigure}{.5\textwidth}
  \centering
  \includegraphics[width=1.0\linewidth]{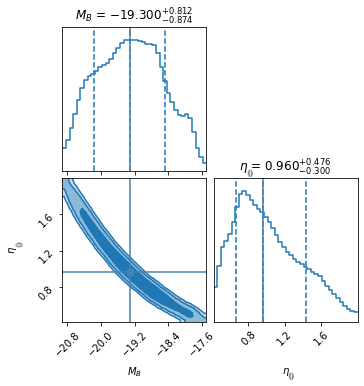}
  \caption{Flat Universe}
  \label{fig:sub1}
\end{subfigure}%
\begin{subfigure}{.5\textwidth}
  \centering
  \includegraphics[width=1.0\linewidth]{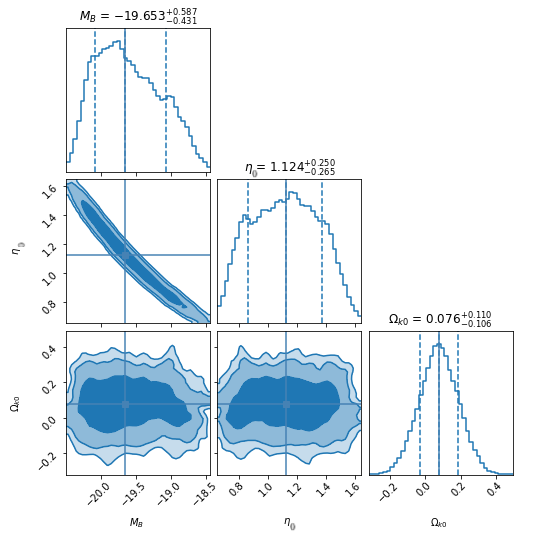}
  \caption{Non-Flat Universe}
  \label{fig:sub2}
\end{subfigure}
\caption{The 1D and 2D posterior distributions of $M_\mathrm{B}, \eta_0$ and $\Omega_{k0}$ for P1 parametrization.}
\label{fig:Part_I_P1}
\end{figure}

\subsubsection*{P2. $\eta(z)=\eta_0+\eta_1 z$.}

In this parametrization, we consider $\eta(z)$ as a function of redshift. The best fit values of $M_\mathrm{B},~\eta_0,~\eta_1$ and $\Omega_{k0}$ are given in Table \ref{tb:table5}.

\begin{table}[h]
	\centering
	\renewcommand{\arraystretch}{2}
	\begin{tabular}[b]{| c | c |c|}\hline
		Parameter & Flat Universe & Non-Flat Universe \\ \hline \hline
		$M_\mathrm{B}$ & $-19.254^{ +0.616}_{ -0.778}$   & $-19.574^{ +0.518}_{ -0.495}$  \\ \hline
		$\eta_0$ & $0.938_{-0.232}^{+0.405}$ & $1.097_{-0.234}^{+0.279}$ \\ \hline
		$\eta_1$ & $0.002^{ +0.010}_{ -0.009}$ & $-0.098_{-0.049}^{+0.047}$ \\ \hline
		$\Omega_{k0}$ & ---  & $1.108_{-0.531}^{+0.503}$ \\ \hline
	\end{tabular}
	\caption{ The best fit values of $M_\mathrm{B},~\eta_0,~\eta_1$ and $\Omega_{k0}$ with $1\sigma$ confidence level for P2 parametrization.}
	\label{tb:table5}
\end{table}

{In case of a flat universe, the best fit value of $\eta_{1}$,  suggests that the cosmic distance duality parameter holds within $1\sigma$ confidence level . Similarly in the case of a non-flat universe, there is no  violation of cosmic distance duality relation at $3\sigma$ confidence level.} The obtained value of $\Omega_{k0}$ can accommodate a flat universe at $3\sigma$ confidence level. The best fit values of $M_B$ in both a flat universe and a non-flat universe remain the same within $1\sigma$ confidence level which indicates that there is no strong impact of $\Omega_{k0}$ on $M_B$.

The 1D and 2D posterior distributions of $M_\mathrm{B}, \eta_0,~\eta_1$ and $\Omega_{k0}$ in both flat and non-flat universes are shown in Fig. \ref{fig:Part_I_P2}. The 2D posterior plots for two cases (flat and non-flat) show a correlation between $M_B$ and distance duality parameters.\\

\begin{figure}
\centering
\begin{subfigure}{.5\textwidth}
  \centering
  \includegraphics[width=1.1\linewidth]{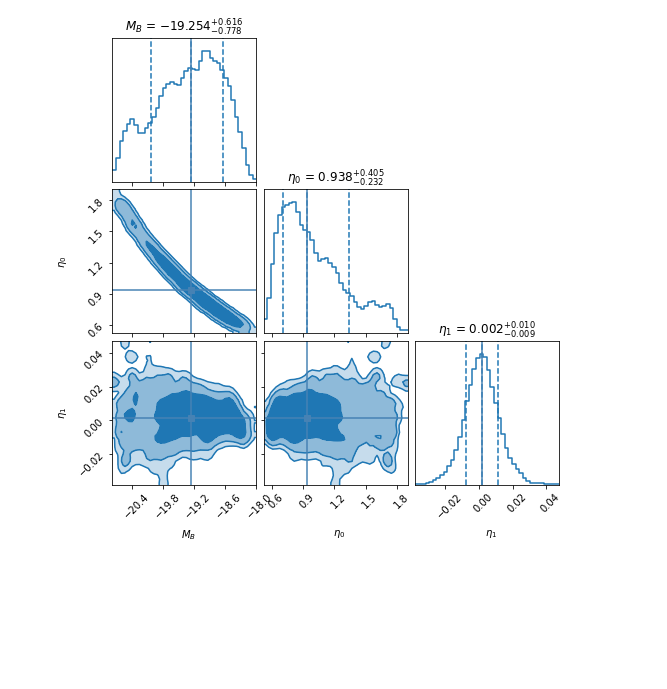}
  \caption{Flat Universe}
  \label{fig:sub1}
\end{subfigure}%
\begin{subfigure}{.5\textwidth}
  \centering
  \includegraphics[width=1.0\linewidth]{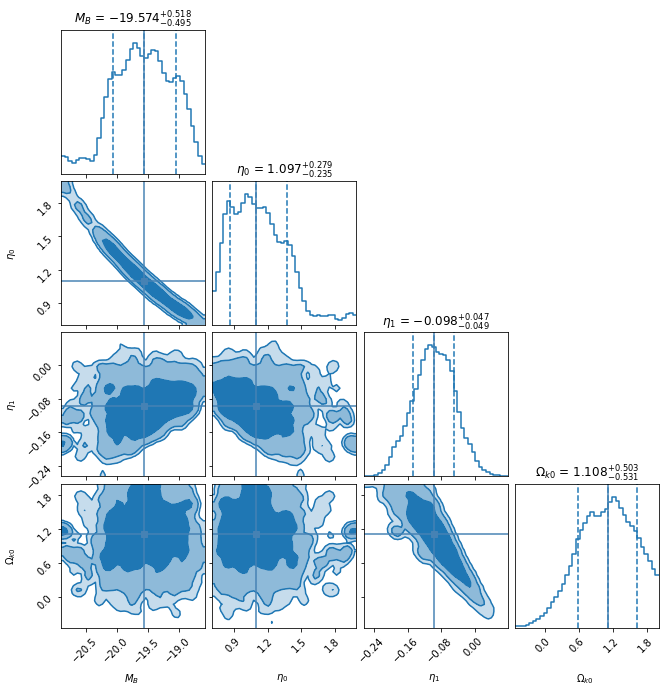}
  \caption{Non-Flat Universe}
  \label{fig:sub2}
\end{subfigure}
\caption{The 1D and 2D posterior distributions of $M_\mathrm{B}, \eta_0,~\eta_1$ and $\Omega_{k0}$ for P2 parametrization.}
\label{fig:Part_I_P2}
\end{figure}

\subsubsection*{P3. $\eta(z)= \eta_0+\eta_1\dfrac{z}{1+z}$.}
In this parametrization, we choose $\eta$ as a function of redshift which converges at high redshift. The best fit values of $M_\mathrm{B},~\eta_0,~\eta_1$ and $\Omega_{k0}$ are given in Table \ref{tb:table7}.\\

{In a flat universe, the CDDR holds within $1\sigma$ confidence level. Even, for the non-flat universe, results indicate that CDDR holds true at $2\sigma$ confidence level. The best fit value of $\Omega_{k0}$ prefer a flat universe at $2\sigma$ confidence level. The best fit values of $M_B$ in both a flat universe and a non-flat universe at $1\sigma$ confidence level indicates that there is no strong correlation between $\Omega_{k0}$ and  $M_B$. }

\begin{table}[H]
	\centering
	\renewcommand{\arraystretch}{2}
	\begin{tabular}[b]{| c | c |c|}\hline
		Parameter & Flat Universe & Non-Flat Universe \\ \hline \hline
		$M_\mathrm{B}$ & $-19.342^{+0.489}_{ -0.369}$   & $-19.368^{ +0.432}_{ -0.335}$ \\ \hline
		$\eta_0$  & $0.976^{ +0.179}_{ -0.197}$ & $0.993^{ +0.166}_{ -0.178}$ \\ \hline
		$\eta_1$  & $0.003^{ +0.015}_{ -0.015}$ & $-0.047^{ +0.036}_{ -0.031}$ \\ \hline
		$\Omega_{k0}$ & --- & $0.312^{ +0.194}_{ -0.204}$    \\ \hline
	\end{tabular}
	\caption{ The best fit values of $M_\mathrm{B},~\eta_0,~\eta_1$ and $\Omega_{k0}$ with $1\sigma$ confidence level for P3 parametrization.}
	\label{tb:table7}
\end{table}

The 1D and 2D posterior distributions of $M_\mathrm{B}, \eta_0,~\eta_1$ and $\Omega_{k0}$ for P3 parametrization of Part I are shown in Fig. \ref{fig:Part_I_P3}.  This figure shows strong correlation between $M_B$ and $\eta_0$,  and $\Omega_{k0}$ and $\eta_1$.\\

\begin{figure}
\centering
\begin{subfigure}{.5\textwidth}
  \centering
  \includegraphics[width=1.0\linewidth]{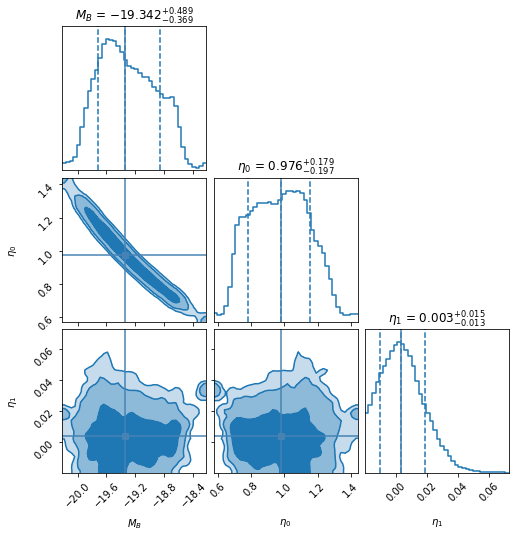}
  \caption{Flat Universe}
  \label{fig:sub1}
\end{subfigure}%
\begin{subfigure}{.5\textwidth}
  \centering
  \includegraphics[width=1.0\linewidth]{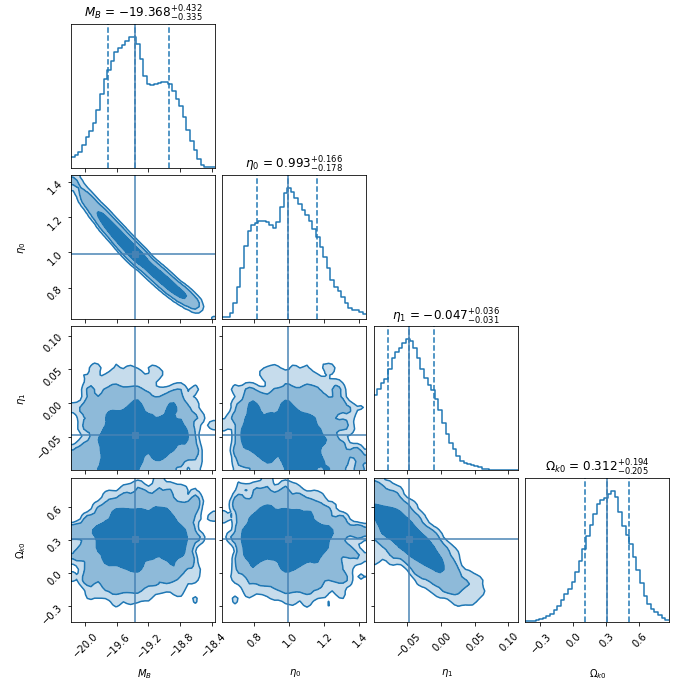}
  \caption{Non-Flat Universe}
  \label{fig:sub2}
\end{subfigure}
\caption{The 1D and 2D posterior distributions of $M_\mathrm{B},~ \eta_0,~\eta_1$ and $\Omega_{k0}$ for P3 parametrization.}
\label{fig:Part_I_P3}
\end{figure}

\subsubsection*{P4. $\eta(z)= \eta_0+\eta_1\text{ln}(1+z)$.}
In this parametrization, we choose $\eta$ as a function of redshift which varies with redshift logarithmic. The best fit value of $M_\mathrm{B},~\eta_0,~\eta_1$ and $\Omega_{k0}$ are given in Table \ref{tb:table7b}.

{In a flat universe, the best fit value of $\eta(z)$ supports the validity of CDDR at $1\sigma$ confidence level. Similarly, the results in a non-flat universe indicate that CDDR does hold true at $2\sigma$ confidence level.} { We find the best fit value of $\Omega_{k0}$ a flat universe at $2\sigma$ confidence level.} The best fit values of $M_B$ in both a flat universe and a non-flat universe indicate that there is no strong correlation between $\Omega_{k0}$ and  $M_B$ at $1 \sigma$ confidence level.

\begin{table}[H]
	\centering
	\renewcommand{\arraystretch}{2}
	\begin{tabular}[b]{| c | c |c|}\hline
		Parameter & Flat Universe & Non-Flat Universe \\ \hline \hline
		$M_\mathrm{B}$ & $-19.202^{ +0.469}_{ -0.505}$   & $-20.000^{ +0.582}_{ -0.402}$ \\ \hline
		$\eta_0$  & $0.916^{ +0.241}_{ -0.178}$ & $1.327^{ +0.265}_{ -0.306}$ \\ \hline
		$\eta_1$  & $0.003^{ +0.014}_{ -0.012}$ & $-0.046^{ +0.041}_{ -0.032}$ \\ \hline
		$\Omega_{k0}$ & --- & $0.333^{+0.215}_{-0.243}$    \\ \hline
	\end{tabular}
	\caption{ The best fit values of $M_\mathrm{B},~\eta_0,~\eta_1$ and $\Omega_{k0}$ with $1\sigma$ confidence level for P4 parametrization.}
	\label{tb:table7b}
\end{table}

The 1D and 2D posterior distributions of $M_\mathrm{B}, \eta_0,~\eta_1$ and $\Omega_{k0}$ for P4 parametrization are shown in Fig. \ref{fig:Part_I_P4}.  This figure shows strong correlation between $M_B$ and $\eta_0$,  and $\Omega_{k0}$ and $\eta_1$.\\

\begin{figure}
\centering
\begin{subfigure}{.5\textwidth}
  \centering
  \includegraphics[width=1.0\linewidth]{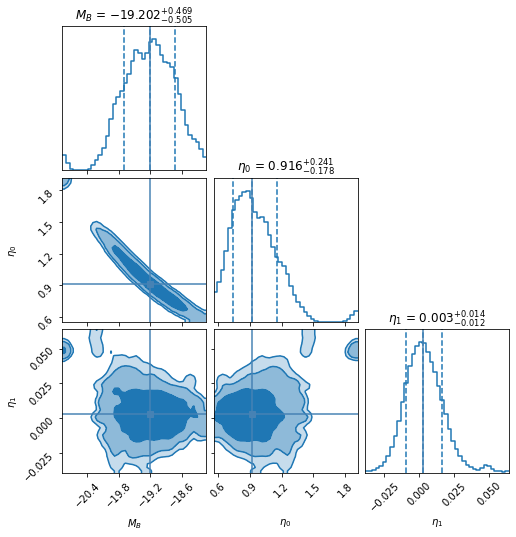}
  \caption{Flat Universe}
  \label{fig:sub1}
\end{subfigure}%
\begin{subfigure}{.5\textwidth}
  \centering
  \includegraphics[width=1.0\linewidth]{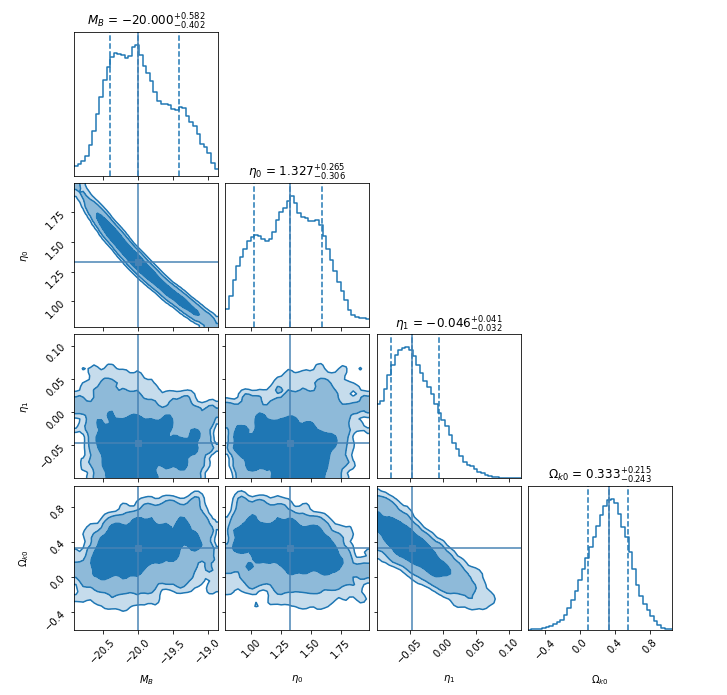}
  \caption{Non-Flat Universe}
  \label{fig:sub2}
\end{subfigure}
\caption{The 1D and 2D posterior distributions of $M_\mathrm{B},~ \eta_0,~\eta_1$ and $\Omega_{k0}$ for P4 parametrization.}
\label{fig:Part_I_P4}
\end{figure}
}

\subsection{Test of variability of SNe Type Ia absolute luminosity}
{In the first part, we find that the cosmic distance duality relation is  not violated at $2\sigma$ confidence level in four  parametrizations of $\eta(z)$ (P1, P2, P3 and P4). 
}

Therefore, in this part we consider that the CDDR is valid and to check the dependency of absolute magnitude on redshift, we  vary $M_B$ with redshift in four different ways in both a flat and a non-flat universe. 
{
\subsubsection*{M1: ${ M_\mathrm{B}(z)=M_\mathrm{B0}}$.}

In the first parametrization of $M_\mathrm{B}$ we consider it as a constant parameter. The best fit values of $M_\mathrm{B}$ and $\Omega_{k0}$ parameters are given in Table \ref{tb:Part_II_P1}.

\begin{table}[h]
	\centering
	\renewcommand{\arraystretch}{2}
	\begin{tabular}[b]{| c | c |c|}\hline
		Parameter & Flat Universe & Non-Flat Universe  \\ \hline \hline
		$M_\mathrm{B0}$ & $-19.390^{ +0.015}_{ -0.015}$   & $-19.393^{ +0.015}_{ -0.015}$ \\ \hline
		$\Omega_{k0}$ & --- & $0.075^{ +0.104}_{ -0.103}$ \\ \hline
		\hline
	\end{tabular}
	\caption{ The best fit values of $M_\mathrm{B0}$ and $\Omega_{k0}$ with $1\sigma$ confidence level for M1 parametrization. }
	\label{tb:Part_II_P1}
\end{table}

{ From table \ref{tb:Part_II_P1}, we find that the best fit value of $M_\mathrm{B}$ is the same in both flat and non-flat universes within $1\sigma$ confidence level. Further, the  best fit value of $\Omega_{k0}=0.075^{ +0.104}_{ -0.103}$ suggests a flat universe  at $1\sigma$ confidence level.} 

The 1D and 2D posterior distributions of $M_\mathrm{B}$ and $\Omega_{k0}$ for M1 parametrization are shown in Fig. \ref{fig:Part_II_P1}.  This figure shows that  there is no correlation between $M_B$ and $\Omega_{k0}$. 

\begin{figure}
\centering
\begin{subfigure}{.5\textwidth}
  \centering
  \includegraphics[width=1.0\linewidth]{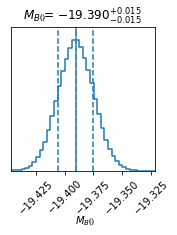}
  \caption{Flat Universe}
  \label{fig:sub1}
\end{subfigure}%
\begin{subfigure}{.5\textwidth}
  \centering
  \includegraphics[width=1.0\linewidth]{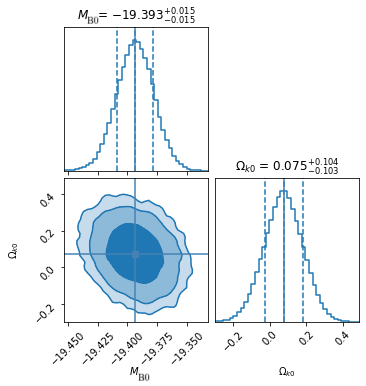}
  \caption{Non-Flat Universe}
  \label{fig:sub2}
\end{subfigure}
\caption{The 1D and 2D posterior distributions of $M_\mathrm{B0}$ and $\Omega_{k0}$ for M1 parametrization.}
\label{fig:Part_II_P1}
\end{figure}


\subsubsection*{M2: ${ M_\mathrm{B}(z)=M_\mathrm{B0}+M_\mathrm{B1} z}$.} 

In this parametrization, we consider $M_\mathrm{B}$ as a function of redshift. The best fit values of $M_\mathrm{B0},~M_\mathrm{B1}$ and $\Omega_{k0}$ are given in Table \ref{tb:Part_II_P2}.

\begin{table}[h]
	\centering
	\renewcommand{\arraystretch}{2}
	\begin{tabular}[b]{| c | c |c|}\hline
		Parameter & Flat Universe & Non-Flat Universe  \\ \hline \hline
		$M_\mathrm{B0}$  & $-19.391^{ +0.016}_{ -0.016}$   & $-19.376^{ +0.018}_{ -0.019}$ \\ \hline
		$M_\mathrm{B1}$  & $0.005^{ +0.021}_{ -0.021}$ & $-0.152^{ +0.089}_{ -0.091}$ \\ \hline
		$\Omega_{k0}$  & ---   & $0.823^{ +0.471}_{ -0.450}$ \\ \hline
	\end{tabular}
	\caption{ The best fit values of $M_\mathrm{B0},~M_\mathrm{B1}$ and $\Omega_{k0}$ with $1\sigma$ confidence level for M2 parametrization. }
	\label{tb:Part_II_P2}
\end{table}

{ In case of a flat universe, we don't find any signal of the redshift evolution of absolute magnitude with $1\sigma$ confidence level. Similarly, in a non-flat universe, there is no redshift dependence of the absolute magnitude at $2\sigma$ confidence level. Further, the result supports a flat universe at  $2\sigma$ confidence level. 
}\\


Figure \ref{fig:Part_II_P2}  illustrates the 1D and 2D posterior distributions of $M_{\mathrm {B0}}, M_{\mathrm {B1}}$ and $\Omega_{k0}$ for M2 parametrization. In this figure, the plot for a non-flat universe indicates a mild correlation between absolute magnitude and cosmic curvature parameter.

\begin{figure}
\centering
\begin{subfigure}{.5\textwidth}
  \centering
  \includegraphics[width=1.0\linewidth]{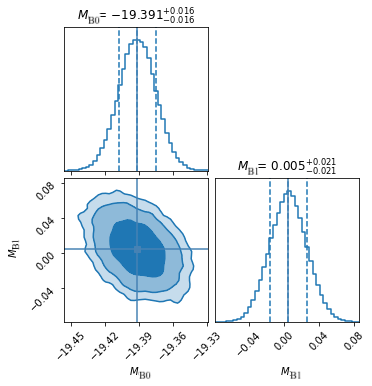}
  \caption{Flat Universe}
  \label{fig:sub1}
\end{subfigure}%
\begin{subfigure}{.5\textwidth}
  \centering
  \includegraphics[width=1.0\linewidth]{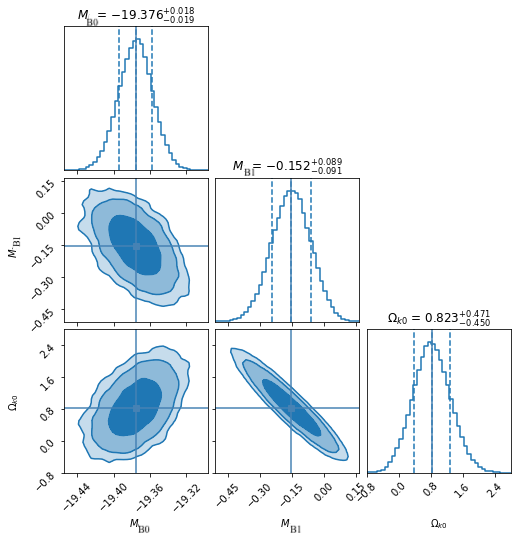}
  \caption{Non-Flat Universe}
  \label{fig:sub2}
\end{subfigure}
\caption{The 1D and 2D posterior distributions of $M_\mathrm{B0}$, $M_\mathrm{B1}$ and $\Omega_{k0}$ for M2 parametrization.}
\label{fig:Part_II_P2}
\end{figure}

\subsubsection*{M3: ${ M_\mathrm{B}(z)=M_\mathrm{B0}+M_\mathrm{B1} \dfrac{z}{1+z}}$.} 

In this parametrization, we choose $M_\mathrm{B}$ as a function of redshift which converges at high redshift. The best fit values of $M_\mathrm{B0},~M_\mathrm{B1}$ and $\Omega_{k0}$ are given in Table \ref{tb:Part_II_P3}.

\begin{table}[h]
	\centering
	\renewcommand{\arraystretch}{2}
	\begin{tabular}[b]{| c | c |c|}\hline
		Parameter & Flat Universe & Non-Flat Universe \\ \hline \hline
		$M_\mathrm{B0}$  & $-19.390^{ +0.017}_{ -0.017}$   & $-19.380^{ +0.018}_{ -0.018}$ \\ \hline
		$M_\mathrm{B1}$  & $0.001^{ +0.038}_{ -0.040}$ & $-0.111^{ +0.082}_{ -0.083}$ \\ \hline
		$\Omega_{k0}$ & ---   & $0.343^{ +0.214}_{ -0.225}$ \\ \hline
	\end{tabular}
	\caption{ The best fit values of $M_\mathrm{B0},~M_\mathrm{B1}$ and $\Omega_{k0}$ with $1\sigma$ confidence level for M3 parametrization. }
	\label{tb:Part_II_P3}
\end{table}

{In this parametrization, the best fit value of $M_\mathrm{B1}$ in both a flat and a non-flat universe suggests that the absolute magnitude does not evolve with redshift at $2\sigma$ confidence level. Furthermore, there is no indication of deviation from flat universe at $2\sigma$ confidence level yet it mildly supports a non flat universe.}\\


\begin{figure}
\centering
\begin{subfigure}{.5\textwidth}
  \centering
  \includegraphics[width=1.0\linewidth]{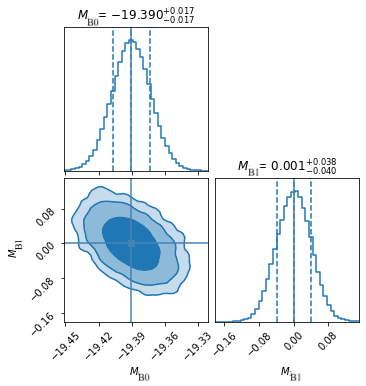}
  \caption{Flat Universe}
  \label{fig:sub1}
\end{subfigure}%
\begin{subfigure}{.5\textwidth}
  \centering
  \includegraphics[width=1.0\linewidth]{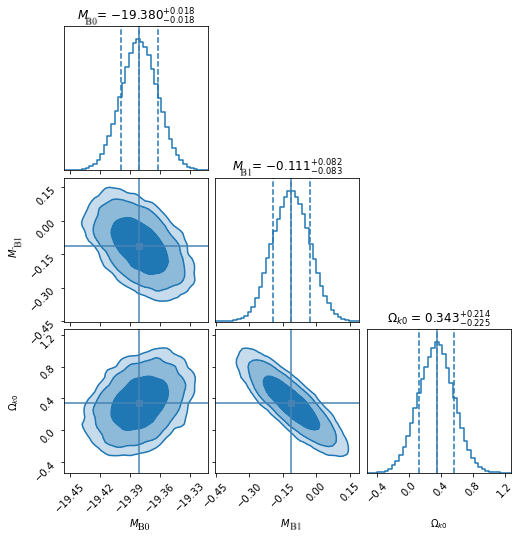}
  \caption{Non-Flat Universe}
  \label{fig:sub2}
\end{subfigure}
\caption{The 1D and 2D posterior distributions of $M_\mathrm{B0}$, $M_\mathrm{B1}$ and $\Omega_{k0}$ for M3 parametrization.}
\label{fig:Part_II_P3}
\end{figure}

In both flat and non-flat universes, the 1D and 2D posterior distributions of $M_\mathrm{B0}$, $M_\mathrm{B1}$ and $\Omega_{k0}$ are illustrated in Figure \ref{fig:Part_II_P3}.  The plot for non-flat universe indicates a correlation between absolute magnitude and cosmic curvature parameter.

\subsubsection*{M4: ${ M_\mathrm{B}(z)=M_\mathrm{B0}+M_\mathrm{B1} \text{ln}(1+z)}$.} 

In this last parametrization, we choose $M_\mathrm{B}$ as a logarithmic function of redshift. The best fit values of $M_\mathrm{B0},~M_\mathrm{B1}$ and $\Omega_{k0}$ are given in Table \ref{tb:Part_II_P4}.

\begin{table}[h]
	\centering
	\renewcommand{\arraystretch}{2}
	\begin{tabular}[b]{| c | c |c|}\hline
		Parameter & Flat Universe & Non-Flat Universe \\ \hline \hline
		$M_\mathrm{B0}$  & $-19.391^{ +0.017}_{ -0.016}$   & $-19.380^{ +0.017}_{ -0.018}$ \\ \hline
		$M_\mathrm{B1}$  & $0.005^{ +0.030}_{ -0.029}$ & $-0.110^{ +0.079}_{ -0.078}$ \\ \hline
		$\Omega_{k0}$ & ---   & $0.442^{ +0.282}_{ -0.287}$ \\ \hline
	\end{tabular}
	\caption{ The best fit values of $M_\mathrm{B0},~M_\mathrm{B1}$ and $\Omega_{k0}$ with $1\sigma$ confidence level for M4 parametrization. }
	\label{tb:Part_II_P4}
\end{table}

In this parametrization, for both a flat and a non-flat universe, the absolute magnitude does not evolve with redshift at $2\sigma$ confidence level. {Furthermore, the best fit value of $\Omega_{k0}$ supports a flat universe at $2\sigma$ confidence level.
}\\
\begin{figure}
\centering
\begin{subfigure}{.5\textwidth}
  \centering
  \includegraphics[width=1.0\linewidth]{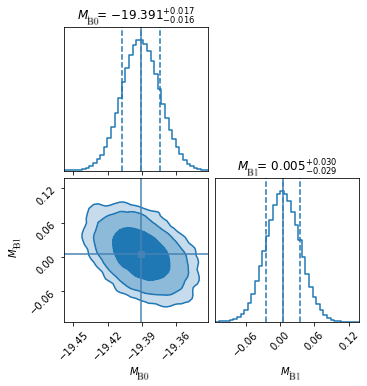}
  \caption{Flat Universe}
  \label{fig:sub1}
\end{subfigure}%
\begin{subfigure}{.5\textwidth}
  \centering
  \includegraphics[width=1.0\linewidth]{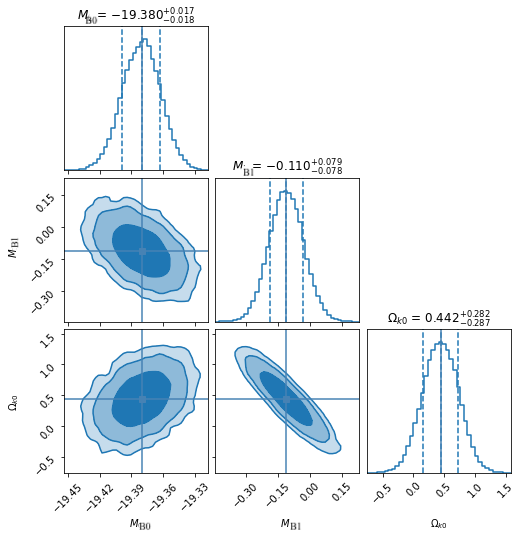}
  \caption{Non-Flat Universe}
  \label{fig:sub2}
\end{subfigure}
\caption{The 1D and 2D posterior distributions of $M_\mathrm{B0}$, $M_\mathrm{B1}$ and $\Omega_{k0}$ for M4 parametrization.}
\label{fig:Part_II_P4}
\end{figure}

In both flat and non-flat universes, the 1D and 2D posterior distributions of $M_\mathrm{B0}$, $M_\mathrm{B1}$ and $\Omega_{k0}$ are illustrated in Figure \ref{fig:Part_II_P4}.  The plot for non-flat universe indicates a correlation between absolute magnitude and cosmic curvature parameter.
}

\section{Discussion and Conclusions}\label{discussion}

In this work, we test the validity of the cosmic distance duality relation which relates the luminosity distance to angular diameter distance via redshift. We use the recent database of $1048$ Type Ia supernovae named Pantheon for the luminosity distance. For the angular diameter distance, we first reconstruct the Hubble parameter, $H(z)$ database of 31 datapoints from Cosmic Chronometers in a model-independent way using the Gaussian Process. Then by adopting the Planck prior on $H_0=67.66 \pm 0.42 \mathrm{~km} \mathrm{~s}^{-1} \mathrm{~Mpc}^{-1}$, the angular diameter distance is obtained corresponding  to the  reconstructed $H(z)$. Further, in the  luminosity distance we have a free parameter, i.e. absolute magnitude of Type Ia supernovae $(M_\mathrm{B})$, and we put constraints on it simultaneously with other cosmological parameters i.e. $\Omega_{k0}$ and $\eta(z)$. Finally, we allow  $M_\mathrm{B}$ and $\eta(z)$ to evolve  with redshift to check whether these are constant quantities or evolve with cosmic time.
\vspace{2mm}\\
We divide  this work into two parts as follows.

\subsection*{\begin{center}
		(A) Part I:  Test of CDDR
\end{center}}
In this part, we do not assume that  cosmic distance duality is  valid and to test this, we consider a free parameter i.e. $\eta(z)$. We fit this parameter simultaneously with $M_\mathrm{B}$ and $\Omega_{k0}$. To check the dependency of the distance duality parameter on redshift, we assume four parametrizations of $\eta(z)$. A brief summary of the results is as follows:
{
\begin{itemize}
	\item {In the case of a flat universe, all parametrizations suggest that the cosmic distance duality relation holds at $1\sigma$ confidence level. 
	}

	\item {For the non-flat universe, P1, supports the  $\eta(z)=1$ at $1\sigma$ confidence level,  P3 \& P4  parametrizations at $2\sigma$ confidence level and,  P2 parametrization at $3\sigma$ confidence level. For this case, this variation in the confidence level for the validity of  $\eta(z)=1$ directly indicates the correlation between the $\eta(z)$ and $\Omega_{k0}$. This analysis also highlights that the results are sensitive to the choice of parametrization and hence, it justifies our use of multiple parametrizations.}

	\item 
Consistently, for flat and non-flat cases, all the 2D contours between $\eta_0$ and $M_\mathrm{B}$ in Part I of the analysis indicate a very strong negative correlation. This points to the need of considering the validity of cosmic distance duality relation in Part II to independently probe the variability of the absolute magnitude $M_\mathrm{B}$.

	\item
		{ In non-flat universe case, the parametrization, P1, strongly suggests a flat universe within $1\sigma$ confidence level while parametrization P2, P3 and P4 also supports the flat universe within $2\sigma$ confidence level. Though the best fit value of $\Omega_{k0}$ mildly prefer a non-flat universe. } \\

\end{itemize}
}

\subsection*{\begin{center}
		(B) Part II: Test of variability of SNe Type Ia absolute luminosity
\end{center}}
{ In the second part we assume  that the cosmic distance duality relation is valid, i.e. $\eta$=1. We are thus left  with two parameters that we have to fit simultaneously. These parameters are $M_\mathrm{B}$ and $\Omega_{k0}$. To test  the dependency of $M_\mathrm{B}$  on redshift, we consider four parametrizations of $M_\mathrm{B}$. All four parametrizations have two model parameters $M_{B0}$ and, $M_{B1}$. While analysing we choose $M_{B0}$ to be constrained in the uniform prior range $U[-21,-17]$ and, $M_{B1}$ in the uniform prior range $U[-2,+2]$. In each parametrization, we discuss two cases. In the first case we consider a flat universe i.e. $\Omega_{k0}=0$ and in the second case we choose a non-flat universe. \\}

{
Our main conclusions of this part are listed below:
\begin{itemize}

    \item  
   { In the flat universe case,  the parametrization $M1$, $M2$ and, $M4$, support no evolution of absolute magnitude with redshift with $1\sigma$ confidence level. Even for the parametrization $M3$, our analysis does not show any signal of redshift evolution of absolute magnitude at $2\sigma$ confidence level.}

    \item 
    { Similarly, in the non-flat universe case, for all parametrizations, no redshift evolution has been found in our analysis  in absolute luminosity $M_{\mathrm{B}}$ at $2\sigma$ confidence level. Hence from our analysis in Part I and Part II, we don't  find any indication towards a statistically significant variability of absolute magnitude $(M_\mathrm{B})$ of type Ia supernova. 
    
    }

    \item {For all parametrizations of the absolute magnitude  $M_{\mathrm{B}}$, the best fit value of $\Omega_{k0}$ suggests a flat universe at $2\sigma$ confidence level. However, in the parametrizations $M2$,$M3$ and M4, the best fit value of $\Omega_{k0}$ show mild preference for a non-flat universe. Further, from the $1D$ and $2D$ contours of all four parametrizations of $M_B(z)$ for non-flat case, we observed a negative correlation between the absolute magnitude and cosmic curvature which should be analysed further.}

\end{itemize}
}

 It is important to note that in  Part I, we fit the distance duality parameter $\eta(z)$ along with $M_B$ and $\Omega_{k0}$. On the other hand, in Part II we assume  the validity of the cosmic distance duality relation i.e; $\eta(z)=1$ and are thus left with  only $M_B~\&~\Omega_{k0}$  as free parameters. Thus, the number of free parameters to be probed decreases from four in Part I to three in Part II. In Part I, our analysis suggests a strong correlation between  the $\eta(z)$ and $M_B$ values, which seems to have an impact on the $1\sigma$ error bars of  $M_B$. However, in Part II, the $\eta(z)$ parameter has been excluded, hence it results in tighter $1\sigma$ constraints on $M_B$. This correlation between $\eta(z)$ and $M_B$ has also been highlighted using BAO and Cluster observations as well \cite{2018lin}.

\subsection*{\begin{center}
		(C) Impact of different prior values of Hubble Parameter ($H_0$)
\end{center}}
    In this analysis, we consider the Planck prior and discuss our results in two parts. In the first part we test the CDDR and in the second part we test the evolution of $M_B$. However, it is important  to check whether our results are sensitive to the chosen prior of $H_0$ or not. To check this sensitivity on the  chosen prior for  $H_0$, we make two more prior choices of $H_0$  while reconstructing the angular diameter distance. These two priors are
    
     \begin{itemize}
     \item 
    No Prior on $H_0$. The reconstructed $H_0$ value using GP is 
    $67.64 \pm 4.79 \mathrm{~km} \mathrm{~s}^{-1} \mathrm{Mpc}^{-1}.$
     \item
     Planck Prior on $H_0=67.66 \pm 0.42 \mathrm{~km} \mathrm{~s}^{-1} \mathrm{Mpc}^{-1}$
     \item
    SH0ES Prior on $H_0=73.20 \pm 1.30 \mathrm{~km} \mathrm{~s}^{-1} \mathrm{Mpc}^{-1}$.
 \end{itemize}

      {To check the impact of the prior in our analysis, we repeat the whole analysis for all parametrizations of $\eta(z)$ and $M_B$ with the three priors of $H_0$ i.e. No prior, Planck prior and SH0ES prior \cite{riess2021}.  In Table \ref{tb:Part_I_P3_Full} and Table \ref{tb:table3full}, we show  the results only for the third parametrization of $\eta(z)$ (i.e $P3$) and $M_B(z)$  ( i.e. $M3$) just to show their behaviour with different priors . The remaining parametrizations in all three priors of $H_0$ give  the same conclusions as we find here in Table \ref{tb:Part_I_P3_Full} and Table \ref{tb:table3full}. 

        \begin{table}[ht]
	\centering
	\renewcommand{\arraystretch}{2}
	\begin{tabular}[b]{| c | c |c| c|}\hline
		Parameter & No Prior & Planck Prior       &SH0ES Prior \\ \hline \hline
		\multicolumn{4}{|c|}{Case.1 Flat Universe}\\ \hline
		$M_\mathrm{B}$ & $-19.161^{ +0.694}_{ -0.590}$ & $-19.342^{ +0.489}_{ -0.369}$   & $-19.038^{ +0.514}_{ -0.454}$ \\ \hline
		$\eta_0$ & $0.927^{ +0.265}_{ -0.254}$ & $0.976^{ +0.179}_{ -0.197}$ & $0.910^{ +0.206}_{ -0.193}$ \\ \hline
		$\eta_1$ & $-0.016^{ +0.008}_{ -0.003}$ & $0.003^{ +0.015}_{ -0.015}$ & $-0.105^{ +0.022}_{ -0.031}$ \\ \hline
		\hline
		\multicolumn{4}{|c|}{Case.2 Non-Flat Universe}\\ \hline
		$M_\mathrm{B}$ & $-19.493^{ +0.433}_{ -0.439}$ & $-19.368^{ +0.432}_{ -0.335}$  & $-19.133^{ +0.563}_{ -0.484}$  \\ \hline
		$\eta_0$ & $1.085^{ +0.256}_{ -0.196}$ & $0.993^{ +0.166}_{ -0.178}$ & $0.948^{ +0.226}_{ -0.217}$ \\ \hline
		$\eta_1$ & $-0.077^{ +0.024}_{ -0.016}$ & $-0.047^{ +0.036}_{ -0.031}$ & $-0.089^{ +0.017}_{ -0.008}$ \\ \hline
		$\Omega_{k0}$ & $0.132^{ +0.152}_{ -0.149}$ & $0.312^{ +0.194}_{ -0.204}$ &  $-0.029^{ +0.135}_{ -0.135}$\\ \hline
	\end{tabular}
	\caption{ The best fit values of $M_\mathrm{B},~\eta_0,~\eta_1$ and $\Omega_{k0}$ with $1\sigma$ confidence level for three $H_0$ priors for P3 parametrization of Part I.}
	\label{tb:Part_I_P3_Full}
\end{table}

\begin{table}[ht]
	\centering
	\renewcommand{\arraystretch}{2}
	\begin{tabular}[b]{| c | c |c| c|}\hline
		Parameter & No Prior & Planck Prior       &SH0ES Prior \\ \hline \hline
		\multicolumn{4}{|c|}{Case.1 Flat Universe}\\ \hline
		$M_\mathrm{B0}$ & $-19.300^{ +0.150}_{ -0.155}$ & $-19.390^{ +0.017}_{ -0.017}$   & $-19.241^{ +0.039}_{ -0.039}$ \\ \hline
		$M_\mathrm{B1}$ & $-0.130^{ +0.040}_{ -0.040}$ & $0.001^{ +0.038}_{ -0.040}$ & $-0.268^{ +0.040}_{ -0.040}$ \\ \hline
		\hline
		\multicolumn{4}{|c|}{Case.2 Non-Flat Universe}\\ \hline
		$M_\mathrm{B0}$ & $-19.286^{ +0.147}_{ -0.167}$ & $-19.380^{ +0.018}_{ -0.018}$  & $-19.229^{ +0.039}_{ -0.039}$  \\ \hline
		$M_\mathrm{B1}$ & $-0.238^{ +0.079}_{ -0.078}$ & $-0.111^{ +0.082}_{ -0.083}$ & $-0.407^{ +0.079}_{ -0.080}$ \\ \hline
		$\Omega_{k0}$ & $0.342^{ +0.213}_{ -0.213}$ & $0.343^{ +0.214}_{ -0.225}$    &  $0.382^{ +0.192}_{ -0.183}$\\ \hline
	\end{tabular}
	\caption{ The best fit values of $M_\mathrm{B0},~M_\mathrm{B1}$ and $\Omega_{k0}$ with $1\sigma$ confidence level for three $H_0$ priors for M3 parametrization of Part II.}
	\label{tb:table3full}
\end{table}

     Through careful analysis, we find that while reconstructing the $H(z)$ data, the choice of the prior value of $H_0$ does not make a significant impact on correlations among the parameters.\\} 
 We also  notice that the $1\sigma$ error bars of the $M_B$ also get affected by the uncertainties of the $H_0$ prior chosen for the analysis. For example, the uncertainty in the value of $H_0$ from CMB is $0.42$ while that  from SH0ES is $1.3$. We find that  in the case of using the  CMB prior on $H_0$, the error bars of $M_B$ are  much  smaller  as compared to  those obtained by using the SH0ES prior. It seems that  the uncertainty in the  prior  propagates in $M_B$ during the fitting analysis and affects the  error bars of $M_B$. Similar to cosmic probes like Cosmic Chronometers and Supernovae Type Ia,  other independent probes like strong gravitational lensing ( Time Delay angular distance measure) have  also seen that the $M_B$ values and error bars are impacted by the associated uncertainties in the input priors, data and parameter values \cite{2020wen}. Given the above mentioned observations, we can  conclude that along with the consideration of the  $\eta(z)$ parameter, the uncertainties in the  prior value of $H_0$  also impacts the $1\sigma$ bounds on the $M_B$. \\

{{Finally, we observe that even if we adopt different priors for $H_0$, the absolute magnitude of SNe Type Ia does not show any redshift dependence in both flat and non-flat universes. This is consistent with the expected non-variability of $M_B$ up to $3\sigma$ confidence level and shows no deviation from the observationally expected assumptions. {{Though our analysis supports a flat universe but we observed the mild preference of best fit value of $\Omega_{k0}$ towards a non flat universe.}}
It should be noted that there is a strong correlation between the absolute magnitude, the distance duality parameter. In Part II, we observed mild correlation between absolute magnitude and cosmic curvature as well.  Any signal of deviation in one of these parameters will impact the rest of the parameters and the underlying assumptions.  We expect that one can resolve  these correlations among different parameters by performing a similar analysis with a larger data set of Supernovae and other observational probes. Upcoming surveys such as the Large Synoptic Survey, the Wide Field Infrared Survey and survey from Large Synoptic Survey Telescope (LSST) 
may assist us in detecting any deviations from the standard supernova type Ia light-curve modeling as well as deviation from the standard cosmological model, i.e. the  $\Lambda$CDM model \cite{2009arXiv0912.0201L, 2019ApJ...873..111I,2018RPPh...81f6901Z,2021MNRAS.507.1746E}.}}

\section*{Acknowledgements}
    We are grateful to the anonymous reviewer for her/his very enlightening remarks which have helped improve the paper.
    Darshan Kumar is supported by an INSPIRE Fellowship under the reference number: IF180293, DST India. A.R. and Darshan Kumar acknowledges facilities provided by the IUCAA Centre for Astronomy Research and Development (ICARD), University of Delhi. In this work some of the figures were created with \textbf{ {\texttt{numpy}}}  \cite{numpy}  and \textbf{{\texttt{matplotlib}}} \cite{matplotlib} Python modules.

\bibliography{references}

\begin{thebibliography}{10}
\newcommand{\enquote}[1]{``#1''}

\bibitem{Riess1998}
A.~G. {Riess} \emph{et~al.}
\newblock {\emph{{Observational Evidence from Supernovae for an Accelerating
  Universe and a Cosmological Constant}}.}
\newblock AJ, \textbf{116}, 1009,  (1998).

\bibitem{Perlmutter1999}
S.~{Perlmutter} \emph{et~al.}
\newblock {\emph{{Measurements of {\ensuremath{\Omega}} and
  {\ensuremath{\Lambda}} from 42 High-Redshift Supernovae}}.}
\newblock ApJ, \textbf{517}, 565,  (1999).

\bibitem{Peebles2003}
P.~J. {Peebles} and B.~{Ratra}.
\newblock {\emph{{The cosmological constant and dark energy}}.}
\newblock Rev. Mod. Phys., \textbf{75}, 559,  (2003).

\bibitem{Caldwell2009}
R.~R. {Caldwell} and M.~{Kamionkowski}.
\newblock {\emph{{The Physics of Cosmic Acceleration}}.}
\newblock Annu. Rev. Nucl. Part. Sci., \textbf{59}, 397,  (2009).

\bibitem{Weinberg2013}
D.~H. {Weinberg}, M.~J. {Mortonson}, D.~J. {Eisenstein}, C.~{Hirata}, A.~G.
  {Riess} and E.~{Rozo}.
\newblock {\emph{{Observational probes of cosmic acceleration}}.}
\newblock Phys. Rep., \textbf{530}, 87,  (2013).

\bibitem{Hillebrandt2000}
W.~{Hillebrandt} and J.~C. {Niemeyer}.
\newblock {\emph{{Type Ia Supernova Explosion Models}}.}
\newblock Annu. Rev. Astron. Astrophys., \textbf{38}, 191,  (2000).

\bibitem{phillips1993}
M.~M. {Phillips}.
\newblock {\emph{{The Absolute Magnitudes of Type IA Supernovae}}.}
\newblock ApJ, \textbf{413}, L105,  (1993).

\bibitem{Tripp1998}
R.~{Tripp}.
\newblock {\emph{{A two-parameter luminosity correction for Type IA
  supernovae}}.}
\newblock A\&A, \textbf{331}, 815,  (1998).

\bibitem{2007ApJ...659..122J}
S.~{Jha}, A.~G. {Riess} and R.~P. {Kirshner}.
\newblock {\emph{{Improved Distances to {Type Ia Supernovae} with Multicolor
  Light-Curve Shapes: MLCS2k2}}.}
\newblock ApJ, \textbf{659}, 122,  (2007).

\bibitem{Salt2}
{J. Guy} \emph{et~al.}
\newblock {\emph{{SALT2:} using distant supernovae to improve the use of type
  {Ia} supernovae as distance indicators}.}
\newblock A\&A, \textbf{466}, 11,  (2007).

\bibitem{Conley_2008}
A.~Conley \emph{et~al.}
\newblock {\emph{{SiFTO}: An Empirical Method for Fitting {SN Ia} Light
  Curves}.}
\newblock ApJ, \textbf{681}, 482–498,  (2008).

\bibitem{Zheng_2017}
W.~Zheng and A.~V. Filippenko.
\newblock {\emph{An Empirical Fitting Method for {Type} {Ia} {Supernova} Light
  Curves: A Case Study of {SN 2011fe}}.}
\newblock ApJ, \textbf{838}, L4,  (2017).

\bibitem{2018ApJ...859..101S}
D.~M. {Scolnic et al.}
\newblock {\emph{{The Complete Light-curve Sample of Spectroscopically
  Confirmed SNe Ia from Pan-STARRS1 and Cosmological Constraints from the
  Combined Pantheon Sample}}.}
\newblock ApJ, \textbf{859}, 101,  (2018).

\bibitem{2013ApJ...770..108C}
M.~{Childress} \emph{et~al.}
\newblock {\emph{{Host Galaxy Properties and Hubble Residuals of Type {Ia}
  Supernovae from the Nearby Supernova Factory}}.}
\newblock ApJ, \textbf{770}, 108,  (2013).

\bibitem{kim_2019}
Y.-L. {Kim}, Y.~{Kang} and Y.-W. {Lee}.
\newblock {\emph{{Environmental Dependence of Type {Ia} Supernova Luminosities
  from the YONSEI Supernova Catalog}}.}
\newblock J. Korean Astron. Soc., \textbf{52}, 181,  (2019).

\bibitem{gallagher2008}
J.~S. {Gallagher}, P.~M. {Garnavich}, N.~{Caldwell}, R.~P. {Kirshner}, S.~W.
  {Jha}, W.~{Li}, M.~{Ganeshalingam} and A.~V. {Filippenko}.
\newblock {\emph{{Supernovae in Early-Type Galaxies: Directly Connecting Age
  and Metallicity with Type Ia Luminosity}}.}
\newblock ApJ, \textbf{685}, 752,  (2008).

\bibitem{2000Hamuy}
M.~{Hamuy}, S.~C. {Trager}, P.~A. {Pinto}, M.~M. {Phillips}, R.~A. {Schommer},
  V.~{Ivanov} and N.~B. {Suntzeff}.
\newblock {\emph{{A Search for Environmental Effects on Type IA Supernovae}}.}
\newblock AJ, \textbf{120}, 1479,  (2000).

\bibitem{wright2018type}
B.~S. {Wright} and B.~{Li}.
\newblock {\emph{{Type Ia supernovae, standardizable candles, and gravity}}.}
\newblock Phys. Rev. D, \textbf{97}, 083505,  (2018).

\bibitem{Drell2000}
P.~S. {Drell}, T.~J. {Loredo} and I.~{Wasserman}.
\newblock {\emph{{Type Ia Supernovae, Evolution, and the Cosmological
  Constant}}.}
\newblock ApJ, \textbf{530}, 593,  (2000).

\bibitem{Combes2004}
F.~{Combes}.
\newblock {\emph{{Properties of SN-host galaxies}}.}
\newblock New Astron. Rev., \textbf{48}, 583,  (2004).

\bibitem{Zehavi1998}
I.~{Zehavi}, A.~G. {Riess}, R.~P. {Kirshner} and A.~{Dekel}.
\newblock {\emph{{A Local Hubble Bubble from Type Ia Supernovae?}}}
\newblock ApJ, \textbf{503}, 483,  (1998).

\bibitem{Conley2007}
A.~{Conley}, R.~G. {Carlberg}, J.~{Guy}, D.~A. {Howell}, S.~{Jha}, A.~G.
  {Riess} and M.~{Sullivan}.
\newblock {\emph{{Is There Evidence for a Hubble Bubble? The Nature of Type Ia
  Supernova Colors and Dust in External Galaxies}}.}
\newblock ApJ, \textbf{664}, L13,  (2007).

\bibitem{Ishak2006}
M.~{Ishak}, A.~{Upadhye} and D.~N. {Spergel}.
\newblock {\emph{{Probing cosmic acceleration beyond the equation of state:
  Distinguishing between dark energy and modified gravity models}}.}
\newblock Phys. Rev. D, \textbf{74}, 043513,  (2006).

\bibitem{Kunz2007}
M.~{Kunz} and D.~{Sapone}.
\newblock {\emph{{Dark Energy versus Modified Gravity}}.}
\newblock Phys. Rev. Lett., \textbf{98}, 121301,  (2007).

\bibitem{Bertschinger2008}
E.~{Bertschinger} and P.~{Zukin}.
\newblock {\emph{{Distinguishing modified gravity from dark energy}}.}
\newblock Phys. Rev. D, \textbf{78}, 024015,  (2008).

\bibitem{Aguirre1999}
A.~{Aguirre}.
\newblock {\emph{{Intergalactic Dust and Observations of Type Ia Supernovae}}.}
\newblock ApJ, \textbf{525}, 583,  (1999).

\bibitem{Rowan2002}
M.~{Rowan-Robinson}.
\newblock {\emph{{Do Type Ia supernovae prove {\ensuremath{\Lambda}}$>$0?}}}
\newblock Mon. Not. Roy. Astron. Soc., \textbf{332}, 352,  (2002).

\bibitem{Goober2002}
A.~{Goobar}, L.~{Bergstr{\"o}m} and E.~{M{\"o}rtsell}.
\newblock {\emph{{Measuring the properties of extragalactic dust and
  implications for the Hubble diagram}}.}
\newblock A\&A, \textbf{384}, 1,  (2002).

\bibitem{Goober2018}
A.~{Goobar}, S.~{Dhawan} and D.~{Scolnic}.
\newblock {\emph{{The cosmic transparency measured with Type Ia supernovae:
  implications for intergalactic dust}}.}
\newblock Mon. Not. Roy. Astron. Soc., \textbf{477}, L75,  (2018).

\bibitem{Avgoustidis2009}
A.~{Avgoustidis}, L.~{Verde} and R.~{Jimenez}.
\newblock {\emph{{Consistency among distance measurements: transparency, BAO
  scale and accelerated expansion}}.}
\newblock JCAP, \textbf{06}, 012,  (2009).

\bibitem{Avgoustidis2010}
A.~{Avgoustidis}, C.~{Burrage}, J.~{Redondo}, L.~{Verde} and R.~{Jimenez}.
\newblock {\emph{{Constraints on cosmic opacity and beyond the standard model
  physics from cosmological distance measurements}}.}
\newblock JCAP, \textbf{08}, 024,  (2010).

\bibitem{Tutusaus2019}
I.~{Tutusaus}, B.~{Lamine} and A.~{Blanchard}.
\newblock {\emph{{Model-independent cosmic acceleration and redshift-dependent
  intrinsic luminosity in type-Ia supernovae}}.}
\newblock A\&A, \textbf{625}, A15,  (2019).

\bibitem{Tutusaus2017}
I.~{Tutusaus}, B.~{Lamine}, A.~{Dupays} and A.~{Blanchard}.
\newblock {\emph{{Is cosmic acceleration proven by local cosmological
  probes?}}}
\newblock A\&A, \textbf{602}, A73,  (2017).

\bibitem{Kang2020}
Y.~{Kang}, Y.-W. {Lee}, Y.-L. {Kim}, C.~{Chung} and C.~H. {Ree}.
\newblock {\emph{{Early-type Host Galaxies of Type Ia Supernovae. II. Evidence
  for Luminosity Evolution in Supernova Cosmology}}.}
\newblock ApJ, \textbf{889}, 8,  (2020).

\bibitem{Valentino_2020}
E.~D. Valentino, S.~Gariazzo, O.~Mena and S.~Vagnozzi.
\newblock {\emph{Soundness of dark energy properties}.}
\newblock JCAP, \textbf{07}, 045,  (2020).

\bibitem{sapone2020measurable}
D.~{Sapone}, S.~{Nesseris} and C.~A.~P. {Bengaly}.
\newblock {\emph{{Is there any measurable redshift dependence on the SN Ia
  absolute magnitude?}}}
\newblock Phys. Dark Universe, \textbf{32}, 100814,  (2021).

\bibitem{Vav2019}
V.~{Vavry{\v{c}}uk}.
\newblock {\emph{{Universe opacity and Type Ia supernova dimming}}.}
\newblock Mon. Not. Roy. Astron. Soc., \textbf{489}, L63,  (2019).

\bibitem{Ellis2007}
G.~F.~R. {Ellis}.
\newblock {\emph{{On the definition of distance in general relativity: I. M. H.
  Etherington (Philosophical Magazine ser. 7, vol. 15, 761 (1933))}}.}
\newblock Gen. Relativ. Gravit., \textbf{39}, 1047,  (2007).

\bibitem{Holanda2010}
R.~F.~L. Holanda, J.~A.~S. Lima and M.~B. Ribeiro.
\newblock {\emph{Testing the distance–duality relation with galaxy clusters
  and type {Ia} supernovae}.}
\newblock ApJ, \textbf{722}, L233,  (2010).

\bibitem{Lima2011}
J.~A.~S. Lima, J.~V. Cunha and V.~T. Zanchin.
\newblock {\emph{Deformed Distance Duality Relations and Supernovae Dimming}.}
\newblock ApJ, \textbf{742}, L26,  (2011).

\bibitem{Li2011}
Z.~Li, P.~Wu and H.~Yu.
\newblock {\emph{Cosmological-model-independent tests for the distance-duality
  relation from Galaxy Clusters and Type {Ia} Supernova}.}
\newblock ApJ, \textbf{729}, L14,  (2011).

\bibitem{Gonalves2011}
R.~S. Gon{\c{c}}alves, R.~F.~L. Holanda and J.~S. Alcaniz.
\newblock {\emph{Testing the cosmic distance duality with X-ray gas mass
  fraction and supernovae data}.}
\newblock Mon. Not. Roy. Astron. Soc., \textbf{420}, L43,  (2011).

\bibitem{Meng2012}
X.-L. Meng, T.-J. Zhang, H.~Zhan and X.~Wang.
\newblock {\emph{Morphology of Galaxy Clusters: A Cosmological
  Model-Independent Test of the Cosmic Distance-Duality Relation}.}
\newblock ApJ, \textbf{745}, 98,  (2012).

\bibitem{Holanda2012}
R.~F.~L. Holanda, R.~S. Gon{\c{c}}alves and J.~S. Alcaniz.
\newblock {\emph{A test for cosmic distance duality}.}
\newblock JCAP, \textbf{06}, 022,  (2012).

\bibitem{Yang2013}
X.~Yang, H.-R. Yu, Z.-S. Zhang and T.-J. Zhang.
\newblock {\emph{{An} {improved} {method} {to} {test} {the}
  {distance}-{duality} {relation}}.}
\newblock ApJ, \textbf{777}, L24,  (2013).

\bibitem{Liang2013}
N.~Liang, Z.~Li, P.~Wu, S.~Cao, K.~Liao and Z.-H. Zhu.
\newblock {\emph{A consistent test of the distance{\textendash}duality relation
  with galaxy clusters and Type {Ia} Supernovae}.}
\newblock Mon. Not. Roy. Astron. Soc., \textbf{436}, 1017,  (2013).

\bibitem{Shafieloo2013}
A.~Shafieloo, S.~Majumdar, V.~Sahni and A.~A. Starobinsky.
\newblock {\emph{Searching for systematics in {SNIa} and galaxy cluster data
  using the cosmic duality relation}.}
\newblock JCAP, \textbf{04}, 042,  (2013).

\bibitem{Zhang:2014eux}
Y.~Zhang.
\newblock {\emph{{Reconstruct the Distance Duality Relation by Gaussian
  Process}}.}
\newblock arXiv:1408.3897,  (2014).

\bibitem{SantosdaCosta2015}
S.~S. da~Costa, V.~C. Busti and R.~F.~L. Holanda.
\newblock {\emph{Two new tests to the distance duality relation with galaxy
  clusters}.}
\newblock JCAP, \textbf{10}, 061,  (2015).

\bibitem{Jhingan2014}
R.~{Nair}, S.~{Jhingan} and D.~{Jain}.
\newblock {\emph{{Observational cosmology and the cosmic distance duality
  relation}}.}
\newblock JCAP, \textbf{05}, 023,  (2011).

\bibitem{Chen2015}
Z.~Chen, B.~Zhou and X.~Fu.
\newblock {\emph{Testing the Distance-Duality Relation from Hubble, Galaxy
  Clusters and Type {Ia} Supernovae Data with Model Independent Methods}.}
\newblock Int. J. Theor. Phys., \textbf{55}, 1229,  (2015).

\bibitem{Holanda2016}
R.~F.~L. Holanda, V.~C. Busti and J.~S. Alcaniz.
\newblock {\emph{Probing the cosmic distance duality with strong gravitational
  lensing and supernovae {Ia} data}.}
\newblock JCAP, \textbf{02}, 054,  (2016).

\bibitem{Rana2016}
A.~Rana, D.~Jain, S.~Mahajan and A.~Mukherjee.
\newblock {\emph{Revisiting the Distance Duality Relation using a
  non-parametric regression method}.}
\newblock JCAP, \textbf{07}, 026,  (2016).

\bibitem{Liao2016}
K.~Liao, Z.~Li, S.~Cao, M.~Biesiada, X.~Zheng and Z.-H. Zhu.
\newblock {\emph{The Distance Duality Relation from Strong Gravitational
  Lensing}.}
\newblock ApJ, \textbf{822}, 74,  (2016).

\bibitem{Holanda2016b}
R.~F.~L. Holanda and K.~N. N.~O. Barros.
\newblock {\emph{Searching for cosmological signatures of the {Einstein}
  equivalence principle breaking}.}
\newblock Phys. Rev. D, \textbf{94}, 023524,  (2016).

\bibitem{Holanda2017}
R.~F.~L. Holanda, S.~H. Pereira and S.~S. da~Costa.
\newblock {\emph{Searching for deviations from the general relativity theory
  with gas mass fraction of galaxy clusters and complementary probes}.}
\newblock Phys. Rev. D, \textbf{95}, 084006,  (2017).

\bibitem{Rana2017}
A.~Rana, D.~Jain, S.~Mahajan, A.~Mukherjee and R.~F.~L. Holanda.
\newblock {\emph{Probing the cosmic distance duality relation using time delay
  lenses}.}
\newblock JCAP, \textbf{07}, 010,  (2017).

\bibitem{Lin2018}
H.-N. Lin, M.-H. Li and X.~Li.
\newblock {\emph{New constraints on the distance duality relation from the
  local data}.}
\newblock Mon. Not. Roy. Astron. Soc., \textbf{480}, 3117,  (2018).

\bibitem{Fu2019}
X.~Fu, L.~Zhou and J.~Chen.
\newblock {\emph{Testing the cosmic distance-duality relation from future
  gravitational wave standard sirens}.}
\newblock Phys. Rev. D, \textbf{99}, 083523,  (2019).

\bibitem{Ruan2018}
C.-Z. Ruan, F.~Melia and T.-J. Zhang.
\newblock {\emph{Model-independent Test of the Cosmic Distance Duality
  Relation}.}
\newblock ApJ, \textbf{866}, 31,  (2018).

\bibitem{Holanda2019}
R.~F.~L. Holanda, L.~R. Cola{\c{c}}o, S.~H. Pereira and R.~Silva.
\newblock {\emph{Galaxy cluster Sunyaev-Zel{\textquotesingle}dovich effect
  scaling-relation and type {Ia} supernova observations as a test for the
  cosmic distance duality relation}.}
\newblock JCAP, \textbf{06}, 008,  (2019).

\bibitem{Chen2020}
J.~Chen.
\newblock {\emph{Testing the distance-duality relation with the baryon acoustic
  oscillations data and type {Ia} supernovae data}.}
\newblock Commun. Theor. Phys., \textbf{72}, 045401,  (2020).

\bibitem{Zheng2020}
X.~Zheng, K.~Liao, M.~Biesiada, S.~Cao, T.-H. Liu and Z.-H. Zhu.
\newblock {\emph{Multiple Measurements of Quasars Acting as Standard Probes:
  Exploring the Cosmic Distance Duality Relation at Higher Redshift}.}
\newblock ApJ, \textbf{892}, 103,  (2020).

\bibitem{Kumar:2020ole}
D.~{Kumar}, D.~{Jain}, S.~{Mahajan}, A.~{Mukherjee} and N.~{Rani}.
\newblock {\emph{{Constraining cosmological and galaxy parameters using strong
  gravitational lensing systems}}.}
\newblock Phys. Rev. D, \textbf{103}, 063511,  (2021).

\bibitem{Hu2018}
J.~Hu and F.~Y. Wang.
\newblock {\emph{Testing the distance{\textendash}duality relation in the
  {Rh~=~ct} universe}.}
\newblock Mon. Not. Roy. Astron. Soc., \textbf{477}, 5064,  (2018).

\bibitem{William2020}
W.~J.~C. {da Silva}, R.~F.~L. {Holanda} and R.~{Silva}.
\newblock {\emph{{Bayesian comparison of the cosmic duality scenarios}}.}
\newblock Phys. Rev. D, \textbf{102}, 063513,  (2020).

\bibitem{2018MNRAS.476.1036M}
J.~{Maga{\~n}a}, M.~H. {Amante}, M.~A. {Garcia-Aspeitia} and V.~{Motta}.
\newblock {\emph{{The Cardassian expansion revisited: constraints from updated
  Hubble parameter measurements and type Ia supernova data}}.}
\newblock Mon. Not. Roy. Astron. Soc., \textbf{476}, 1036,  (2018).

\bibitem{Jimenez_2002}
R.~Jimenez and A.~Loeb.
\newblock {\emph{Constraining Cosmological Parameters Based on Relative Galaxy
  Ages}.}
\newblock ApJ, \textbf{573}, 37,  (2002).

\bibitem{Moresco_2012}
M.~Moresco \emph{et~al.}
\newblock {\emph{Improved constraints on the expansion rate of the Universe up
  to z$\sim$ 1.1 from the spectroscopic evolution of cosmic chronometers}.}
\newblock JCAP, \textbf{08}, 006,  (2012).

\bibitem{Zhang_2014}
C.~Zhang, H.~Zhang, S.~Yuan, S.~Liu, T.-J. Zhang and Y.-C. Sun.
\newblock {\emph{Four new {observational H}(z) data from luminous red galaxies
  in the Sloan Digital Sky Survey data release seven}.}
\newblock Res. Astron. Astrophys., \textbf{14}, 1221,  (2014).

\bibitem{PhysRevD.71.123001}
J.~Simon, L.~Verde and R.~Jimenez.
\newblock {\emph{Constraints on the redshift dependence of the dark energy
  potential}.}
\newblock Phys. Rev. D, \textbf{71}, 123001,  (2005).

\bibitem{Stern_2010}
D.~Stern, R.~Jimenez, L.~Verde, M.~Kamionkowski and S.~A. Stanford.
\newblock {\emph{Cosmic chronometers: constraining the equation of state of
  dark energy. {I:H(z)} measurements}.}
\newblock JCAP, \textbf{02}, 008,  (2010).

\bibitem{10.1093/mnrasl/slv037}
M.~Moresco.
\newblock {\emph{{Raising the bar: new constraints on the Hubble parameter with
  cosmic chronometers at z $\sim$ 2}}.}
\newblock Mon. Not. Roy. Astron. Soc., \textbf{450}, L16,  (2015).

\bibitem{Moresco_2016}
M.~Moresco \emph{et~al.}
\newblock {\emph{A 6{\%} measurement of the Hubble parameter at z$\sim$0.45:
  direct evidence of the epoch of cosmic re-acceleration}.}
\newblock JCAP, \textbf{05}, 014,  (2016).

\bibitem{Planck2018}
{Planck Collaboration} \emph{et~al.}
\newblock {\emph{{Planck 2018 results. VI. Cosmological parameters}}.}
\newblock A\&A, \textbf{641}, A6,  (2020).

\bibitem{2006gpml}
C.~E. Rasmussen.
\newblock {\emph{Gaussian processes in machine learning}.}
\newblock Summer school on machine learning, Springer,  (2003).

\bibitem{2020PhRvD.102b3520K}
L.~{Kazantzidis} and L.~{Perivolaropoulos}.
\newblock {\emph{{Hints of a local matter underdensity or modified gravity in
  the low z Pantheon data}}.}
\newblock Phys. Rev. D, \textbf{102}, 023520,  (2020).

\bibitem{2001IJMPD..10..213C}
M.~{Chevallier} and D.~{Polarski}.
\newblock {\emph{{Accelerating Universes with Scaling Dark Matter}}.}
\newblock Int. J. Mod. Phys. D, \textbf{10}, 213,  (2001).

\bibitem{2003PhRvL..90i1301L}
E.~V. {Linder}.
\newblock {\emph{{Exploring the Expansion History of the Universe}}.}
\newblock Phys. Rev. Lett., \textbf{90}, 091301,  (2003).

\bibitem{2005MNRAS.356L..11J}
H.~K. {Jassal}, J.~S. {Bagla} and T.~{Padmanabhan}.
\newblock {\emph{{WMAP constraints on low redshift evolution of dark energy}}.}
\newblock Mon. Not. Roy. Astron. Soc., \textbf{356}, L11,  (2005).

\bibitem{2013PASP..125..306F}
D.~{Foreman-Mackey}, D.~W. {Hogg}, D.~{Lang} and J.~{Goodman}.
\newblock {\emph{{emcee: The MCMC Hammer}}.}
\newblock Publ. Astron. Soc. Pac., \textbf{125}, 306,  (2013).

\bibitem{2016JOSS....1...24F}
D.~{Foreman-Mackey}.
\newblock {\emph{{corner.py: Scatterplot matrices in Python}}.}
\newblock J. Open Source Softw., \textbf{1}, 24,  (2016).

\bibitem{2018lin}
H.-N. Lin, M.-H. Li and X.~Li.
\newblock {\emph{New constraints on the distance duality relation from the
  local data}.}
\newblock Monthly Notices of the Royal Astronomical Society, \textbf{480},
  3117–3122,  (2018).

\bibitem{riess2021}
A.~G. Riess, S.~Casertano, W.~Yuan, J.~B. Bowers, L.~Macri, J.~C. Zinn and
  D.~Scolnic.
\newblock {\emph{Cosmic Distances Calibrated to 1\% Precision with Gaia EDR3
  Parallaxes and Hubble Space Telescope Photometry of 75 Milky Way Cepheids
  Confirm Tension with $\Lambda$CDM}.}
\newblock ApJ, \textbf{908}, L6,  (2021).

\bibitem{2020wen}
X.~Wen and K.~Liao.
\newblock {\emph{Calibrating the standard candles with strong lensing}.}
\newblock The European Physical Journal C, \textbf{80},  (2020).

\bibitem{2009arXiv0912.0201L}
{LSST Science Collaboration} \emph{et~al.}
\newblock {\emph{{LSST Science Book, Version 2.0}}.}
\newblock arXiv e-prints, arXiv:0912.0201,  (2009).

\bibitem{2019ApJ...873..111I}
{\v{Z}}.~{Ivezi{\'c}} \emph{et~al.}
\newblock {\emph{{LSST: From Science Drivers to Reference Design and
  Anticipated Data Products}}.}
\newblock ApJ, \textbf{873}, 111,  (2019).

\bibitem{2018RPPh...81f6901Z}
H.~{Zhan} and J.~A. {Tyson}.
\newblock {\emph{{Cosmology with the Large Synoptic Survey Telescope: an
  overview}}.}
\newblock Rep. Prog. Phys., \textbf{81}, 066901,  (2018).

\bibitem{2021MNRAS.507.1746E}
T.~{Eifler} \emph{et~al.}
\newblock {\emph{{Cosmology with the Roman Space Telescope - multiprobe
  strategies}}.}
\newblock Mon. Not. Roy. Astron. Soc., \textbf{507}, 1746,  (2021).

\bibitem{numpy}
T.~E. Oliphant.
\newblock \emph{A guide to NumPy}, vol.~1,  (Trelgol Publishing USA2006).

\bibitem{matplotlib}
J.~D. Hunter.
\newblock {\emph{Matplotlib: A 2D graphics environment}.}
\newblock Comput. Sci. Eng., \textbf{9}, 90,  (2007).

\end{thebibliography}
\bibliographystyle{Darshan_Custom_ref} 

\end{document}